\documentclass[11pt]{article}

\usepackage[final]{acl}
\usepackage{amssymb}
\usepackage{amsmath}
\usepackage{multirow}
\usepackage{booktabs}
\usepackage{url}

\usepackage{times}
\usepackage{latexsym}

\usepackage[T1]{fontenc}
\usepackage[utf8]{inputenc}

\usepackage{microtype}

\usepackage{inconsolata}

\usepackage{graphicx}

\title{MGRetrieval: Memory-Guided Reflective Retrieval for\\ Long-Term Dialogue Agents}

\author{Tan Wang \and Yunwei Dong \\
Northwestern Polytechnical University \\
Xi'an, Shaanxi, China \\
\texttt{330754522@nwpu.edu.cn} \and \texttt{}
}

\begin{document}
\maketitle
\begin{abstract}
Large Language Models (LLMs) have made significant progress in dialogue, yet redundant memory contexts severely limit their effectiveness in long-term dialogue agents. 
External memory systems have been proposed to improve memory maintenance.
However, these systems mainly rely on one-shot retrieval, which limits their ability to retrieve sufficient and relevant evidence.
Although recent methods introduce reflection into retrieval, their retrieval paths are generated by the LLM from limited evidence, leading to unstable retrieval and additional latency overhead.
%These limitations highlight the need for effective retrieval mechanisms.
To address these limitations, we propose MGRetrieval, a retrieval strategy that grounds reflective retrieval in the semantic structure of historical memories.
Specifically, MGRetrieval consists of two steps:
(1) It references the structure of historical memories to construct a more precise retrieval path.
(2) The LLM retains critical memories and determines whether accumulated memories are sufficient to stop further iterative retrieval.
This allows the retrieval process to follow semantically meaningful paths.
Through memory-guided retrieval and critical memory propagation, MGRetrieval gradually constructs concise and sufficient memory contexts. 
Extensive experiments on LoCoMo show that MGRetrieval outperforms the strongest baseline by 8.91\% in F1 and 11.11\% in BLEU-1 on average across Qwen2.5-14B and Qwen3-14B, while maintaining practical token and latency costs.
The code can be found in \url{https://anonymous.4open.science/r/MGRetrieval}.

\end{abstract}

\section{Introduction}
\begin{figure*}[t]
    \centering
    \includegraphics[width=\textwidth]{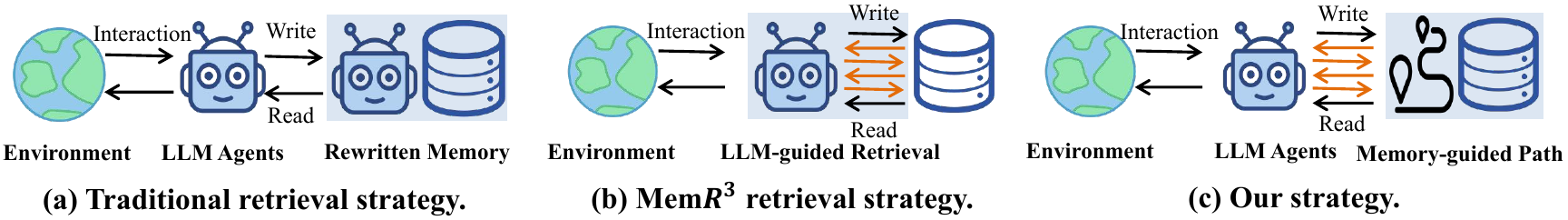}
    \caption{(a) Traditional retrieval strategies perform a one-shot retrieval step. (b) Reflective retrieval strategies, such as MemR$^3$, rely on the LLM to guide the iterative retrieval. (c) Our strategy constructs a local memory-guided retrieval path and uses the LLM to preserve critical memory and determine when to stop.}
    \label{fig:introduction}
\end{figure*}
Large language models (LLMs) have demonstrated remarkable dialogue capabilities~\cite{mendoncca2024benchmarking}. 
However, in long-term dialogue, a large portion of extended histories may be irrelevant or noisy for the current query~\cite{liu2024lost,pan2025memory}.
Such redundant memory contexts can further hinder effective reasoning and incur substantial token overhead~\cite{liu2024lost,li2023compressing}.
%Supporting long-term dialogue often requires recall relevant information from extended histories~\cite{maharana2024evaluating,wu2024longmemeval}.
As a result, constructing concise and sufficient memory contexts remains a significant challenge for long-term dialogue agents.

External memory systems integrating various memory management mechanisms represent a significant step to improving memory maintenance.
MemoryBank~\cite{zhong2024memorybank} maintains long-term user memories with a forgetting-inspired update mechanism, while A-Mem~\cite{xu2025mem} organizes interactions into evolving notes.
%MemoryBank \cite{zhong2024memorybank} maintains a long-term memory with an updating mechanism inspired by the Ebbinghaus forgetting curve, while 
%A-Mem \cite{xu2025mem} builds a structured memory system that organizes interactions into evolving notes.
MemoryOS \cite{kang2025memory} establishes a comprehensive memory operating system by coordinating memory storage, updating, retrieval, and generation.
%These systems improve memory maintenance, but their retrieval often remains one-shot. 
However, these systems retrieve memories one-shot for a query without explicitly checking whether the retrieved evidence is sufficient. 
As a result, their memory contexts may fail to retrieve sufficient evidence or introduce irrelevant memories~\cite{zhang2026evoking}.
Reflective retrieval methods attempt to address this limitation by retrieving evidence iteratively. 
MemR$^3$~\cite{du2025memr} uses the LLM to preserve critical memories, decide whether to continue retrieval and refine the next retrieval query.
%It avoids introducing redundant memory through iterative small-scope retrieval.
However, its next retrieval query is generated by the LLM from limited retrieved memories, without explicitly considering the broader structure of the memory bank.
This may make retrieval coarse or misaligned with the needed memory regions, leading to suboptimal memory contexts and additional latency overhead.
These limitations highlight retrieval control as an underexplored aspect in agent memory management.

To address these limitations, we propose MGRetrieval, a memory-guided reflective retrieval strategy for long-term dialogue agents, as shown in Figure~\ref{fig:introduction}.
MGRetrieval integrates two core mechanisms:
\textbf{Memory-guided Retrieval} extracts keywords as compact semantic abstractions of individual memories.
It then constructs a keyword pyramid to order retrieval from specific to broader semantic regions.
This guides the retrieval process to consistently target regions of the memory bank with clear semantic support.
\textbf{Memory Filtering} leverages the LLM to retain critical memories as a summary for current retrieval, while filtering out memories retrieved in previous turns.
This maintains concise memory contexts and reduces token and latency costs without losing relevant evidence.
The LLM leverages accumulated evidence to determine whether to continue retrieval.
Through memory-guided iterative retrieval and critical memory propagation, MGRetrieval builds concise and semantically sufficient memory contexts. 
MGRetrieval surpasses the strongest baseline on LoCoMo across diverse response generation metrics.

Our contributions are summarized as follows:
\begin{enumerate}
    \item We introduce MGRetrieval, a memory-guided reflective retrieval strategy that constructs a retrieval path from historical memories to build concise and sufficient memory contexts for long-term dialogue agents.
    \item MGRetrieval derives a keyword pyramid from the query and historical memories, progressively expands retrieval scopes, and uses the LLM to identify critical memories and assess evidence sufficiency.
    \item Experimental results show that MGRetrieval improves long-term conversational memory reasoning capabilities of LLM agents while maintaining practical token and latency costs.
\end{enumerate}

\section{Related Work}
%\subsection{Memory for LLM Agents}
%Large language models (LLMs) face challenges in long-term agent tasks because only a small part of historical memory is relevant to the current query~\cite{liu2024lost,wu2024longmemeval}. 
%Simply extending the input context may introduce irrelevant information, thereby increase computation cost, and even weaken reasoning~\cite{li2023compressing,xiao2024duoattention}. 
%Therefore, LLM agents should construct concise and task-relevant memory contexts before reasoning~\cite{pan2025memory,cao2024retaining}.

\subsection{Memory System}
Existing studies develop dedicated memory systems for LLM agents to enhance their long-term memory management capabilities.
They primarily cover memory storage, forgetting, updating, and retrieval mechanisms~\cite{zhang2025survey}.
%Existing studies develop dedicated mechanisms to build memory systems for LLM agents, enhancing their long-term interaction capabilities~\cite{zhang2025survey}.
For example, MemoryBank \cite{zhong2024memorybank} maintains long-term user memories with an updating mechanism inspired by the Ebbinghaus forgetting curve and retrieves memories via dense retrieval with FAISS indexing.
A-Mem \cite{xu2025mem} stores interactions as atomic and semantically linked notes, allowing memories to evolve through dynamic indexing and linking, and retrieves relevant notes based on similarity.
%Memory-R1~\cite{yan2025memory} uses reinforcement learning to improve memory management and utilization in LLM agents.
MemoryOS \cite{kang2025memory} introduces a hierarchical framework to organize, update, and retrieve memories at different levels.
These methods mainly focus on organizing and updating memories to maintain compact and useful memory banks.
However, their retrieval strategies remain largely one-shot, retrieving memories once without verifying their relevance.
MGRetrieval is a retrieval-centric mechanism leveraging reflection to perform iterative retrieval.

\subsection{Reflective Retrieval}
Recent studies~\cite{jiang2023active,trivedi2023interleaving,shao2023enhancing} show that one-shot retrieval can be insufficient for complex information needs, especially when the required evidence is indirect, long-form, or multi-hop.
Self-RAG~\cite{asai2024self} introduces self-reflection into RAG to adaptively decide when to retrieve and how to use retrieved evidence, thereby improving the controllability and quality of generation.
%It is not designed specifically for long-term agent memory, but they show that retrieval should be controlled adaptively rather than performed only once.
RMM~\cite{tan2025prospect} studies reflective memory management for long-term LLM agents, using prospective and retrospective reflection to organize and retrieve relevant memories.
MemR$^3$ \cite{du2025memr} further introduces reflection-based closed-loop control into memory retrieval and shows that iterative reflection can improve retrieval.
Its router selects among retrieval, reflection, and answering actions based on the task and current evidence.
During this loop, the system uses the LLM to identify remaining evidence gaps and generate the next retrieval query.
In contrast, MGRetrieval uses the memory bank structure for precise retrieval, rather than relying on the LLM and limited evidence as in other methods.

\section{MGRetrieval}
\begin{figure*}[t]
    \centering
    \includegraphics[width=\textwidth]{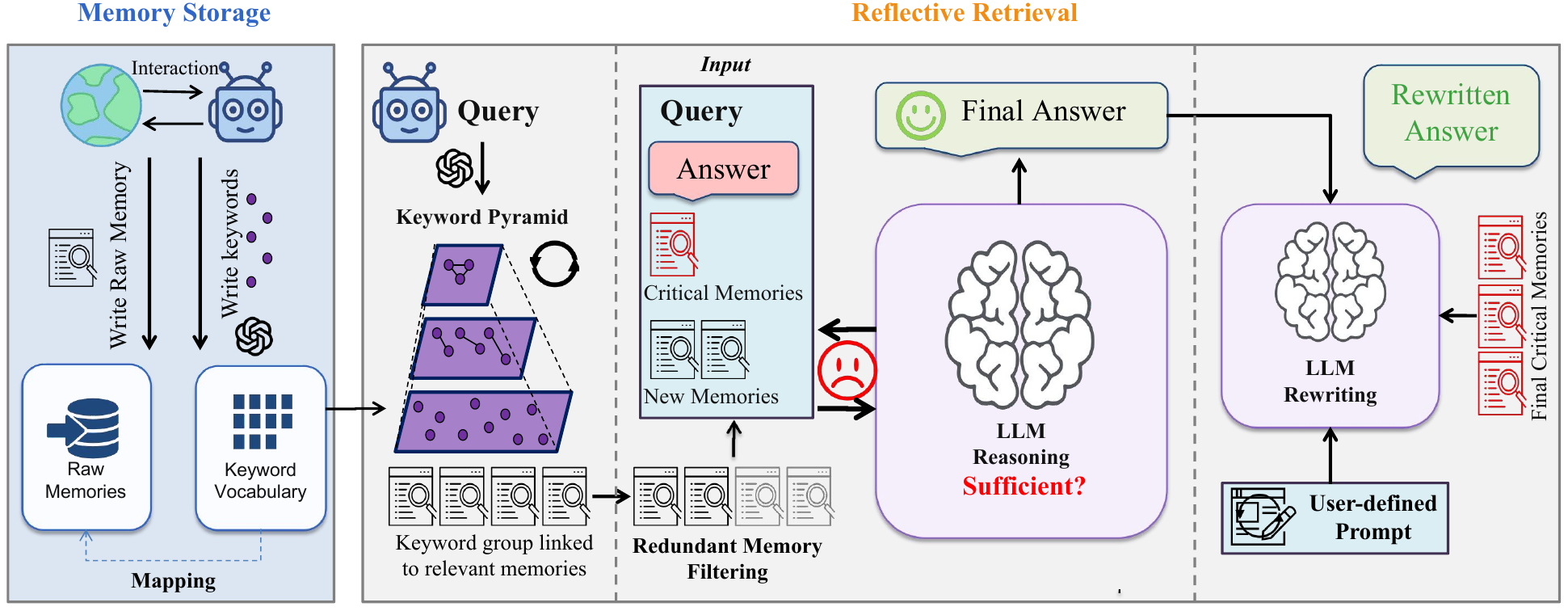}
    \caption{The overall architecture of MGRetrieval, including memory storage, retrieval pyramid construction, redundant memory filtering, LLM-based assessment and answering, and final answer rewriting.}
    \label{fig:model}
\end{figure*}
{MGRetrieval} is a memory-guided retrieval strategy for long-term dialogue agents. 
The core idea is precisely retrieval along a memory-guided path while accumulating critical memories through reflection until sufficient evidence is available.

Specifically, MGRetrieval consists of a memory storage module and a reflective retrieval module, as shown in Figure~\ref{fig:model}. The retrieval module comprises four mechanisms: keyword retrieval pyramid construction, redundant memory filtering, LLM assessment and answering, and answer rewriting.

\subsection{Memory Storage}
The storage module does not compress or overwrite original memories.
{MGRetrieval} extracts keywords from each memory to construct a keyword vocabulary and builds mappings between each keyword and its associated memories. 
It treats keywords as compact semantic abstractions of individual memories. 
The keyword vocabulary provides a lightweight semantic index over the memory bank, and the mapping preserves the association between semantic cues and original interactions.
This index is later used to construct localized retrieval paths in the retrieval pyramid.

{MGRetrieval} extracts keywords from each historical memory $m_i=(q_i,a_i)$ using the LLM with a keyword extraction prompt $p_{s1}$:
\begin{equation}
K_i = \mathcal{LLM}(m_i; p_{s1}).    
\end{equation}
To reduce vocabulary redundancy and link new memories to existing vocabulary entries, MGRetrieval uses the LLM with a vocabulary matching prompt \(p_{s2}\):
\begin{equation}
T_i = \mathcal{C}_{\mathrm{LLM}}(K_i, \mathcal{V}_{i-1}; p_{s2}),
\end{equation}
where \(T_i\) contains the extracted keywords after vocabulary matching.
This enables each memory to be indexed by both newly extracted keywords and matched vocabulary entries.
The new keywords are collected:
\begin{equation}
\mathcal{N}_i
=
\left\{\, k \in K_i \mid k\notin\mathcal{V}_{i-1} \,\right\}.
\end{equation}
The vocabulary is then updated by inserting new keywords:
\begin{equation}
\mathcal{V}_i
=
\mathcal{V}_{i-1}
\cup
\mathcal{N}_i .
\end{equation}

{MGRetrieval} preserves all original memories and establishes explicit mappings. 
For each keyword, the newly associated memories are defined as:
\begin{equation}
\Delta_i(v)
=
\{\,m_i \mid v \in T_i \,\}.
\end{equation}
The mapping is then updated as:
\begin{equation}
\Phi_i(v)
=
\begin{cases}
\Phi_{i-1}(v)\cup \Delta_i(v), & v\in \mathcal{V}_{i-1},\\
\Delta_i(v), & v\in \mathcal{N}_i.
\end{cases}
\end{equation}
Here, $\Phi_i(v)$ denotes the set of memories associated with keyword $v$ after adding memory $m_i$.

\subsection{Memory Retrieval}
MGRetrieval first constructs a memory-guided control path, implemented as a keyword pyramid.
It then retrieves new memories along the pyramid and combines them with the critical memories from previous rounds, the query, and the previous answer to form the current input.
The LLM answers the question and assesses whether additional retrieval is required.
Finally, it rewrites the output to produce a personalized and standardized answer.
\paragraph{Retrieval Pyramid Construction.} 
%{MGRetrieval} adopts keyword-based retrieval to avoid complex computational overhead, making it better suited to the practical requirement of timely response.
MGRetrieval selects query-relevant keywords from vocabulary and organizes them into a keyword pyramid, which defines the order of retrieval scope expansion.

For a given query \(q\), let \(\mathcal{V}\) denote the existing keyword vocabulary and \(p_{r1}\) denote the query keyword selection prompt.
MGRetrieval uses the LLM to select a query-relevant keyword set \(Q\):
\begin{equation}
Q = \mathcal{LLM}(q,\mathcal{V};p_{r1}), \quad Q\subseteq \mathcal{V}. 
\end{equation}
Then it constructs a retrieval pyramid $P$ based on the keyword set \(Q=\{k_1,k_2,\dots,k_n\}\). 
Let ${P}^{(l)}_j$ denote the $j$-th keyword group at level $l$:
\begin{equation}
{P}^{(l)}_j=\{k_{i_1},\dots,k_{i_l}\}, \quad 1\le i_1<\cdots<i_l\le n, 
\end{equation}
\begin{equation}
{P}^{(l)}=\{{P}^{(l)}_j\}_{j=1}^{\binom{n}{l}}, \quad l=1,\dots,n.
\end{equation}
Each level \(l\) enumerates all \(l\)-keyword combinations from \(Q\).
The lowest level \(l=1\) contains individual keywords, while higher levels contain keyword groups composed of more keywords. 
For each keyword group, we retrieve its associated memories as:
\begin{equation}
s(P^{(l)}_j)
=
\bigcap_{v\in P^{(l)}_j}\Phi(v).
\end{equation}
Higher-level keyword groups retrieve more semantically specific memories, but they are also more susceptible to over-specialization.
At the same level, the keyword groups are ranked in descending order by the number of associated memories and different groups cover different semantic aspects.
MGRetrieval traverses the pyramid from the top level \(l=n\) to the bottom level \(l=1\), progressively expanding the semantic scope of retrieval while collecting evidence from different semantic perspectives at each level.
%In contrast, lower-level keyword groups tend to retrieve more memories, including those already covered at higher levels.
%This design enables MGRetrieval to perform efficient retrieval with a progressively expanding scope along the pyramid.

\paragraph{Memory Filtering.} To keep each input concise, MGRetrieval retains only new memories retrieved in this round and asks the LLM to return critical memories as a summary of retrieved memories.
%It preserves the critical memories from previous rounds and combines them with newly retrieved memories to form the current memory input.

At round \(r\), let \(P^{(l_r)}_{j_r}\) denote the currently selected keyword group, and \(R^{(r)} = s(P^{(l_r)}_{j_r})\) denote the current retrieved memories.

The effective memory input in the current round, denoted as $\tilde{R}^{(r)}$, excludes memories $\bigcup_{t=1}^{r-1} R^{(t)}$ that have already appeared in previous rounds:
\begin{equation}
\tilde{R}^{(r)}
=
R^{(r)} \setminus \bigcup_{t=1}^{r-1} R^{(t)}.
\end{equation}
The critical memories are returned in the answer and assessment step.
%Meanwhile, it incorporates the critical memories $H^{(r-1)}$ obtained in the last round $r-1$ to support the reasoning. 
%$H^{(r-1)}$ serves as a compact summary of prior reasoning, allowing the model to preserve important information without reintroducing redundant inputs.
\paragraph{Answer and Assessment.}  
In each round, the LLM adaptively assesses whether the current memories are sufficient to answer.
If not, {MGRetrieval} preserves the critical memories and retrieves from the next keyword group to collect additional relevant memories. 
This iterative process continues until the accumulated evidence is judged sufficient to answer the query, the maximum number of iterations is reached, or the pyramid is exhausted.

The newly retrieved memories \(\tilde{R}^{(r)}\) are integrated with the previously identified critical memories \(H^{(r-1)}\), the previous answer \(a^{(r-1)}\), and the original query \(q\). 
We denote the input at round \(r\) as
\begin{equation}
I^{(r)}=\left(q, a^{(r-1)}, H^{(r-1)}, \tilde{R}^{(r)}\right).
\end{equation}
The reasoning is formulated with prompt \(p_{r2}\) as
\begin{equation}
a^{(r)}, b^{(r)}, H^{(r)}
=
\mathcal{LLM}\!\left(I^{(r)}; p_{r2}\right).
\end{equation}
Here, $\mathcal{LLM}$ denotes the LLM call used for reasoning at round \(r\).
The model outputs the current answer $a^{(r)}$, a binary flag $b^{(r)}$ indicating whether the answer should be accepted, and the critical memories $H^{(r)}$ identified at this round. 

\paragraph{Answer Rewriting.} To normalize the final answer format, MGRetrieval uses a rewriting module.
This module does not retrieve additional memory. 
It refines the final answer using only the final critical memories and a user-defined prompt:
\begin{equation}
\hat{a}=\mathcal{LLM}\!\left(a,\; H;\; p_{r3}\right),
\end{equation}
where $a$ denotes the final answer, $H$ denotes the final critical memory set, $p_{r3}$ is the prompt that specifies the user-defined formatting requirements, and $\hat{a}$ denotes the answer after rewriting.
\section{Experiment}
\begin{table}[t]
\centering
\caption{Weighted average results on LoCoMo using Qwen2.5-14B and Qwen3-14B.}
\label{tab:metrics}

\resizebox{\columnwidth}{!}{%
\begingroup
\fontsize{9}{12}\selectfont
\renewcommand{\arraystretch}{1.1}
\setlength{\tabcolsep}{1.2pt}

\begin{tabular}{c|ccccc}
\hline
Method & F1 & BLEU-1 & ROUGE-L & ROUGE-2 & METEOR \\
\hline
MemoryBank   & 19.56 & 16.52 & 30.36 & 21.72 & 29.62 \\
FULL       & \underline{39.27} & 32.92 & 33.85 & 20.02 & 29.14 \\
RAG          & 34.62 & 30.45 & 34.28 & 19.21 & 28.31 \\
A-Mem        & 32.43 & 23.30 & 33.93 & 18.48 & 23.10 \\
MemR$^3$     & 35.30 & 30.46 & 35.36 & \underline{22.47} & \underline{30.89} \\
MemoryOS     & 38.48 & \underline{33.02} & \underline{37.04} & 21.04 & 30.38 \\
MGRetrieval & \textbf{42.77} & \textbf{36.69} & \textbf{41.62} & \textbf{24.22} & \textbf{35.56} \\
\hline
Improvement  & +8.91\% & +11.11\% & +12.37\% & +7.79\% & +15.12\% \\
\hline
\end{tabular}

\endgroup
}
\end{table}

\begin{table}[t]
\centering
\fontsize{12}{15}\selectfont
\caption{Comparison results on the GVD dataset.}
\label{tab:gvd}

\begin{tabular}{c|ccc}
\hline
Method & Acc. & Corr. & Cohe. \\
\hline
MemoryBank   & 79.0 & 74.5 & 90.5 \\
RAG          & 92.0 & 92.0 & 92.0 \\
A-Mem        & 91.0 & 89.5 & 91.5 \\
MemR$^3$     & 93.0 & 92.0 & 92.5 \\
MemoryOS     & 93.0 & 91.0 & 92.0 \\
MGRetrieval & 93.0 & 91.5 & 92.5 \\
\hline
\end{tabular}
\end{table}

\begin{table}[t]
\centering
\fontsize{8}{10}\selectfont
\caption{Average main LLM cost per memory reconstruction and per question response. The auxiliary LLM cost in MGRetrieval is described in Section~\ref{sec:efficiency}.}
\label{tab:cost}
\begin{tabular}{lccccc}
\hline
\multirow{2}{*}{Model} & \multicolumn{2}{c}{Memory} & \multicolumn{3}{c}{Response} \\
\cline{2-3} \cline{4-6}
&  Calls &  Tokens &  Calls &  Tokens & Time (s) \\
\hline
MemoryBank   & 0.29 & 275     & 1     & 1177  & 3.27  \\
FULL         & 0    & 0       & 1     & 21332 & 2.56  \\
RAG          & 0    & 0       & 1     & 2091  & 3.31\\
A-Mem        & 3.05 & 1832    & 2     & 3975  & 3.52 \\
MemR$^3$     & 0    & 0       & 3.27  & 5169  & 12.77\\
MemoryOS     & 4.41 & 1268    & 2     & 4537  & 2.77\\
MGRetrieval & 0    & 0       & 2.23  & 5799  & 6.98 \\
\hline
\end{tabular}
\end{table}

\begin{table*}[tb!]
\centering
\caption{
Detailed F1 and BLEU-1 results on the LoCoMo dataset across four categories (Multi Hop, Temporal, Open Domain, and Single Hop).
Average is recomputed as a weighted average over the four categories.
}
\label{tab:main}
\resizebox{\textwidth}{!}{%
\begin{tabular}{ccl|cccccccc|cc}
\hline
\multicolumn{2}{c}{\multirow{3}{*}{\textbf{Model}}} & \multicolumn{1}{c|}{\multirow{3}{*}{\textbf{Method}}} & \multicolumn{8}{c|}{\textbf{Category}} & \multicolumn{2}{c}{\textbf{}} \\ \cline{4-13}
\multicolumn{2}{c}{} & \multicolumn{1}{c|}{} 
& \multicolumn{2}{c|}{\textbf{Multi Hop}} 
& \multicolumn{2}{c|}{\textbf{Temporal}} 
& \multicolumn{2}{c|}{\textbf{Open Domain}} 
& \multicolumn{2}{c|}{\textbf{Single Hop}} 
& \multicolumn{2}{c}{\textbf{Average}} \\
\multicolumn{2}{c}{} & \multicolumn{1}{c|}{} 
& \textbf{F1} & \multicolumn{1}{c|}{\textbf{BLEU}} 
& \textbf{F1} & \multicolumn{1}{c|}{\textbf{BLEU}} 
& \textbf{F1} & \multicolumn{1}{c|}{\textbf{BLEU}} 
& \textbf{F1} & \multicolumn{1}{c|}{\textbf{BLEU}} 
& \textbf{F1} & \textbf{BLEU} \\ \hline

\multirow{14}{*}{\textbf{\rotatebox{90}{GPT}}} 
& \multicolumn{1}{c|}{\multirow{7}{*}{\textbf{\rotatebox{90}{4o-mini}}}} 
& \textsc{MemoryBank} 
& 5.00 & \multicolumn{1}{c|}{4.77} 
& 9.68 & \multicolumn{1}{c|}{6.99} 
& 5.56 & \multicolumn{1}{c|}{5.94} 
& 6.61 & \multicolumn{1}{c|}{5.16} 
& 6.89 & 5.52 \\

& \multicolumn{1}{c|}{} 
& \textsc{FULL} 
& 25.02 & \multicolumn{1}{c|}{19.75} 
& 18.41 & \multicolumn{1}{c|}{14.77} 
& 12.04 & \multicolumn{1}{c|}{11.16} 
& 40.36 & \multicolumn{1}{c|}{29.05} 
& 31.21 & 23.26 \\

& \multicolumn{1}{c|}{} 
& \textsc{RAG} 
& 19.59 & \multicolumn{1}{c|}{15.84} 
& 36.11 & \multicolumn{1}{c|}{31.96} 
& 13.03 & \multicolumn{1}{c|}{9.69} 
& 43.60 & \multicolumn{1}{c|}{39.87} 
& 35.74 & 31.94 \\

& \multicolumn{1}{c|}{} 
& \textsc{A-Mem} 
& 27.02 & \multicolumn{1}{c|}{20.09} 
& 45.85 & \multicolumn{1}{c|}{36.67} 
& 12.14 & \multicolumn{1}{c|}{12.00} 
& 44.65 & \multicolumn{1}{c|}{37.06} 
& 39.65 & 32.31 \\

& \multicolumn{1}{c|}{} 
& \textsc{MemR$^3$} 
& 29.27 & \multicolumn{1}{c|}{22.48} 
& 36.18 & \multicolumn{1}{c|}{32.62} 
& 20.11 & \multicolumn{1}{c|}{16.11} 
& \textbf{53.00} & \multicolumn{1}{c|}{\textbf{48.09}} 
& 43.03 & 38.11 \\

& \multicolumn{1}{c|}{} 
& \textsc{MemoryOS} 
& \textbf{35.20} & \multicolumn{1}{c|}{\textbf{26.55}} 
& 43.33 & \multicolumn{1}{c|}{38.02} 
& \textbf{25.89} & \multicolumn{1}{c|}{\textbf{19.24}} 
& 48.79 & \multicolumn{1}{c|}{43.85} 
& 43.74 & 37.93 \\

& \multicolumn{1}{c|}{} 
& \textsc{MGRetrieval} 
& 31.95 & \multicolumn{1}{c|}{22.20} 
& \textbf{49.72} & \multicolumn{1}{c|}{\textbf{44.15}} 
& 23.70 & \multicolumn{1}{c|}{18.42} 
& 52.18 & \multicolumn{1}{c|}{46.94} 
& \textbf{46.19} & \textbf{40.05} \\ \cline{2-13}

& \multicolumn{1}{c|}{\multirow{7}{*}{\textbf{\rotatebox{90}{4o}}}} 
& \textsc{MemoryBank} 
& 6.49 & \multicolumn{1}{c|}{4.69} 
& 2.47 & \multicolumn{1}{c|}{2.43} 
& 6.43 & \multicolumn{1}{c|}{5.30} 
& 8.28 & \multicolumn{1}{c|}{7.10} 
& 6.63 & 5.57 \\

& \multicolumn{1}{c|}{} 
& \textsc{FULL} 
& 28.00 & \multicolumn{1}{c|}{18.47} 
& 9.09 & \multicolumn{1}{c|}{5.78} 
& 16.47 & \multicolumn{1}{c|}{14.80} 
& \textbf{61.56} & \multicolumn{1}{c|}{\textbf{54.19}} 
& 41.67 & 35.10 \\

& \multicolumn{1}{c|}{} 
& \textsc{RAG} 
& 23.23 & \multicolumn{1}{c|}{19.08} 
& 48.86 & \multicolumn{1}{c|}{44.99} 
& 13.79 & \multicolumn{1}{c|}{11.76} 
& 44.68 & \multicolumn{1}{c|}{40.85} 
& 39.70 & 35.91 \\

& \multicolumn{1}{c|}{} 
& \textsc{A-Mem} 
& 32.86 & \multicolumn{1}{c|}{23.76} 
& 39.41 & \multicolumn{1}{c|}{31.23} 
& 17.10 & \multicolumn{1}{c|}{15.84} 
& 48.43 & \multicolumn{1}{c|}{42.97} 
& 41.75 & 35.31 \\

& \multicolumn{1}{c|}{} 
& \textsc{MemR$^3$} 
& 35.12 & \multicolumn{1}{c|}{26.63} 
& 40.28 & \multicolumn{1}{c|}{35.27} 
& 23.11 & \multicolumn{1}{c|}{18.71} 
& 54.91 & \multicolumn{1}{c|}{49.56} 
& 46.19 & 40.39 \\

& \multicolumn{1}{c|}{} 
& \textsc{MemoryOS} 
& \textbf{41.30} & \multicolumn{1}{c|}{\textbf{30.70}} 
& 44.24 & \multicolumn{1}{c|}{37.71} 
& 25.96 & \multicolumn{1}{c|}{20.33} 
& 47.84 & \multicolumn{1}{c|}{41.53} 
& 44.53 & 37.43 \\

& \multicolumn{1}{c|}{} 
& \textsc{MGRetrieval} 
& 33.12 & \multicolumn{1}{c|}{24.01} 
& \textbf{51.24} & \multicolumn{1}{c|}{\textbf{46.37}} 
& \textbf{26.88} & \multicolumn{1}{c|}{\textbf{22.75}} 
& 52.48 & \multicolumn{1}{c|}{45.72} 
& \textbf{47.08} & \textbf{40.45} \\ \hline

\multirow{14}{*}{\textbf{\rotatebox{90}{Qwen}}} 
& \multicolumn{1}{c|}{\multirow{7}{*}{\textbf{\rotatebox{90}{2.5-14B}}}} 
& \textsc{MemoryBank} 
& 7.64 & \multicolumn{1}{c|}{5.70} 
& 1.93 & \multicolumn{1}{c|}{1.45} 
& 1.63 & \multicolumn{1}{c|}{1.28} 
& 20.23 & \multicolumn{1}{c|}{16.19} 
& 12.95 & 10.27 \\

& \multicolumn{1}{c|}{} 
& \textsc{FULL} 
& 31.60 & \multicolumn{1}{c|}{23.81} 
& 32.93 & \multicolumn{1}{c|}{25.47} 
& 16.00 & \multicolumn{1}{c|}{13.30} 
& 42.84 & \multicolumn{1}{c|}{36.72} 
& 37.04 & 30.55 \\

& \multicolumn{1}{c|}{} 
& \textsc{RAG} 
& 18.94 & \multicolumn{1}{c|}{15.25} 
& 25.74 & \multicolumn{1}{c|}{21.31} 
& 14.05 & \multicolumn{1}{c|}{10.83} 
& 41.20 & \multicolumn{1}{c|}{36.81} 
& 32.21 & 28.02 \\

& \multicolumn{1}{c|}{} 
& \textsc{A-Mem} 
& 25.27 & \multicolumn{1}{c|}{18.37} 
& 23.51 & \multicolumn{1}{c|}{18.02} 
& 13.39 & \multicolumn{1}{c|}{12.35} 
& 37.30 & \multicolumn{1}{c|}{32.20} 
& 30.73 & 25.47 \\

& \multicolumn{1}{c|}{} 
& \textsc{MemR$^3$} 
& 31.12 & \multicolumn{1}{c|}{23.39} 
& 11.20 & \multicolumn{1}{c|}{8.89} 
& 18.66 & \multicolumn{1}{c|}{15.38} 
& 49.91 & \multicolumn{1}{c|}{43.67} 
& 36.35 & 30.85 \\

& \multicolumn{1}{c|}{} 
& \textsc{MemoryOS} 
& \textbf{34.68} & \multicolumn{1}{c|}{\textbf{26.32}} 
& 30.03 & \multicolumn{1}{c|}{24.85} 
& 23.93 & \multicolumn{1}{c|}{21.01} 
& 44.77 & \multicolumn{1}{c|}{39.89} 
& 38.55 & 33.09 \\

& \multicolumn{1}{c|}{} 
& \textsc{MGRetrieval} 
& 32.37 & \multicolumn{1}{c|}{22.54} 
& \textbf{35.29} & \multicolumn{1}{c|}{\textbf{30.80}} 
& \textbf{27.77} & \multicolumn{1}{c|}{\textbf{22.08}} 
& \textbf{52.36} & \multicolumn{1}{c|}{\textbf{46.22}} 
& \textbf{43.61} & \textbf{37.16} \\ \cline{2-13}

& \multicolumn{1}{c|}{\multirow{7}{*}{\textbf{\rotatebox{90}{3-14B}}}} 
& \textsc{MemoryBank} 
& 16.15 & \multicolumn{1}{c|}{12.23} 
& 13.22 & \multicolumn{1}{c|}{10.64} 
& 6.41 & \multicolumn{1}{c|}{5.56} 
& 36.73 & \multicolumn{1}{c|}{32.89} 
& 26.17 & 22.77 \\

& \multicolumn{1}{c|}{} 
& \textsc{FULL} 
& 32.88 & \multicolumn{1}{c|}{23.70} 
& 35.70 & \multicolumn{1}{c|}{30.14} 
& 14.35 & \multicolumn{1}{c|}{10.71} 
& 49.71 & \multicolumn{1}{c|}{43.97} 
& 41.50 & 35.30 \\

& \multicolumn{1}{c|}{} 
& \textsc{RAG} 
& 21.11 & \multicolumn{1}{c|}{16.94} 
& 35.96 & \multicolumn{1}{c|}{30.75} 
& 11.96 & \multicolumn{1}{c|}{9.66} 
& 45.63 & \multicolumn{1}{c|}{41.71} 
& 37.04 & 32.89 \\

& \multicolumn{1}{c|}{} 
& \textsc{A-Mem} 
& 29.56 & \multicolumn{1}{c|}{13.77} 
& 24.02 & \multicolumn{1}{c|}{9.61} 
& 12.02 & \multicolumn{1}{c|}{11.23} 
& 42.04 & \multicolumn{1}{c|}{29.10} 
& 34.13 & 21.12 \\

& \multicolumn{1}{c|}{} 
& \textsc{MemR$^3$} 
& 28.31 & \multicolumn{1}{c|}{21.68} 
& 19.70 & \multicolumn{1}{c|}{17.11} 
& 13.23 & \multicolumn{1}{c|}{10.20} 
& 44.34 & \multicolumn{1}{c|}{40.25} 
& 34.26 & 30.08 \\

& \multicolumn{1}{c|}{} 
& \textsc{MemoryOS} 
& \textbf{34.61} & \multicolumn{1}{c|}{\textbf{26.39}} 
& 32.41 & \multicolumn{1}{c|}{27.06} 
& 18.81 & \multicolumn{1}{c|}{14.83} 
& 44.22 & \multicolumn{1}{c|}{39.44} 
& 38.41 & 32.94 \\

& \multicolumn{1}{c|}{} 
& \textsc{MGRetrieval} 
& 29.56 & \multicolumn{1}{c|}{21.49} 
& \textbf{36.35} & \multicolumn{1}{c|}{\textbf{30.99}} 
& \textbf{23.75} & \multicolumn{1}{c|}{\textbf{19.04}} 
& \textbf{50.27} & \multicolumn{1}{c|}{\textbf{45.13}} 
& \textbf{41.92} & \textbf{36.23} \\ \hline

\end{tabular}%
}
\end{table*}
\subsection{Experimental Settings}
\paragraph{Datasets.} We evaluate MGRetrieval on conversational memory benchmarks.
We use the LoCoMo dataset \cite{maharana2024evaluating}, which contains substantially longer multi-session conversations. 
It is particularly suitable for comprehensively evaluating the memory abilities of LLMs over extended conversations.
We also conduct experiments on the GVD \cite{zhong2024memorybank} dataset to evaluate the effectiveness of our memory strategy. 
Detailed dataset descriptions are provided in Appendix~\ref{Datasets}.

\paragraph{Metrics.} For LoCoMo, we adopt five evaluation metrics: F1, which measures the overall balance between precision and recall, BLEU-1 \cite{papineni2002bleu}, which evaluates the word-level overlap, and additional metrics (ROUGE-L, ROUGE-2, and METEOR). For GVD, we report memory retrieval accuracy (Acc.), response correctness (Corr.), and contextual coherence (Cohe.). 
Detailed definitions can be found in Appendix~\ref{Metrics}.

\paragraph{Compared Baselines.} We compare MGRetrieval with five representative methods, including comprehensive memory systems such as MemoryBank  \cite{zhong2024memorybank}, A-Mem \cite{xu2025mem}, and MemoryOS \cite{kang2025memory}, as well as retrieval-centric methods such as RAG \cite{karpukhin2020dense} and MemR$^3$ \cite{du2025memr}.
Detailed baseline descriptions are provided in Appendix~\ref{Baseline}.

\paragraph{Implementation Details.} 
Throughout all experiments, we fix the number of pyramid layers to 4 and set the maximum number of iterations to 4.
We consistently use the lightweight GPT-4o-mini to assist keyword extraction, while different LLMs are used only for memory bank construction and answer generation.
The rewriting prompt draws on the LoCoMo guidelines and includes a few examples.
Details can be found in Appendix~\ref{Implementation Details}.
\subsection{Main Results}
We evaluate MGRetrieval in terms of answer quality, memory context length, and practical token and latency costs to assess its effectiveness and efficiency.
Details can be found in Appendix~\ref{Results}.
\paragraph{Performance Analysis.}
The performance results on the LoCoMo and GVD benchmarks are reported in Tables \ref{tab:metrics}, \ref{tab:main} and \ref{tab:gvd}.

On LoCoMo, FULL directly uses the complete memory as input. We make the following observations:
(1) Memory systems such as MemoryOS achieve competitive or strong performance compared with FULL.
This highlights the value of external memory modules in long-term dialogue agents.
(2) However, MemoryBank and A-Mem obtain lower scores than retrieval-centric approaches such as RAG and MemR$^3$ in some cases.
Moreover, MemR$^3$ achieves better performance than many one-shot methods such as RAG and A-Mem by using reflective retrieval.
This comparison highlights the importance of retrieval and suggests that reflective retrieval is a more effective retrieval strategy for long-term memory management.
(3) In contrast, MGRetrieval achieves higher average scores than MemR$^3$ which employs an LLM-guided reflective retrieval strategy and the overall strongest baseline MemoryOS, across the five evaluation metrics.
Even in the relatively weaker categories, its results still remain competitive. 
For instance, averaged over Qwen2.5-14B and Qwen3-14B, MGRetrieval outperforms the second-best result by 8.91\% in F1 and 11.11\% in BLEU-1.
These results indicate that memory-guided paths enable more targeted retrieval of relevant memories, thereby improving agent reasoning in long-term dialogue.

On the GVD dataset, which contains shorter dialogues and simpler scenarios, MGRetrieval achieves performance comparable to the strongest baselines. With most metrics near saturation, these results mainly show that MGRetrieval maintains robustness across different scenarios.

\begin{figure}[t]
    \centering
    \includegraphics[width=0.95\columnwidth]{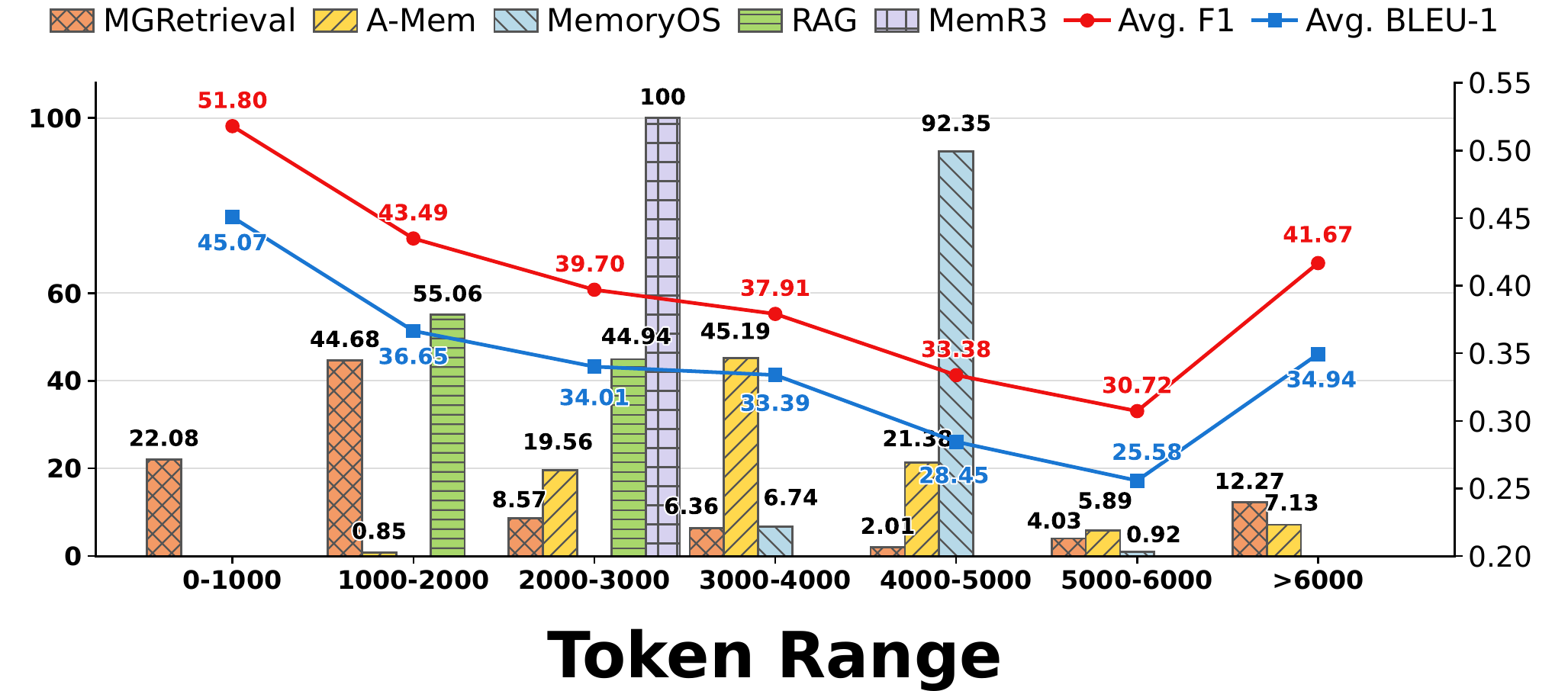}
    \caption{Final memory context lengths across different methods and the performance of MGRetrieval under different length intervals on Qwen2.5-14B.}
    \label{fig:token}
\end{figure}

\paragraph{Input Analysis.}
Figure \ref{fig:token} shows the distribution of the memory context lengths.
We observe that other methods tend to produce longer memory contexts, whereas MGRetrieval keeps memory contexts concise while achieving stronger performance.
MGRetrieval also exhibits a broader length distribution and performs better in low-token regimes.
This pattern suggests that MGRetrieval adaptively adjusts its memory contexts according to query difficulty.
It tends to construct inputs with highly useful memories rather than introducing large amounts of context at once, as other methods often do.
For complex questions, MGRetrieval adaptively expands the retrieval scope to gather additional information.
We attribute this to memory-guided paths, which exploit the organization of the memory bank to prioritize highly relevant memories, and to the filtering of redundant contexts.
%{MGRetrieval} produces concise yet effective final inputs and thereby enhances the LLMs' ability to reason.
As a result, MGRetrieval can regulate context length across questions with different retrieval demands.
%Its strong performance further indicates that the core of memory management lies in efficient retrieval: Precise retrieval reduces memory redundancy and ensures memory relevance, thereby enhancing the reasoning ability of LLM agents.

\begin{table*}[t]
\centering
\fontsize{10}{12}\selectfont
\renewcommand{\arraystretch}{1.2}
\setlength{\tabcolsep}{3pt}
\caption{Ablation results on LoCoMo with Qwen2.5-14B. Round 1 (w/o R) also corresponds to a one-shot variant before memory-guided iterative expansion.}
\label{tab:ablation}
\resizebox{\textwidth}{!}{%
\begin{tabular}{l|ccccccc|ccccccc}
\hline
\multicolumn{1}{c|}{\multirow{3}{*}{\textbf{Stage}}}
& \multicolumn{7}{c|}{\textbf{(w/o RMR)}}
& \multicolumn{7}{c}{\textbf{Base}} \\
\cline{2-15}
& \multicolumn{1}{c|}{\multirow{2}{*}{\textbf{Samples}}}
& \multicolumn{3}{c|}{\textbf{Current}}
& \multicolumn{3}{c|}{\textbf{Full}}
& \multicolumn{1}{c|}{\multirow{2}{*}{\textbf{Samples}}}
& \multicolumn{3}{c|}{\textbf{Current}}
& \multicolumn{3}{c}{\textbf{Full}} \\
\cline{3-8} \cline{10-15}
& \multicolumn{1}{c|}{}
& \textbf{F1} & \textbf{BLEU-1} & \multicolumn{1}{c|}{\textbf{Tokens}}
& \textbf{F1} & \textbf{BLEU-1} & \multicolumn{1}{c|}{\textbf{Tokens}}
& \multicolumn{1}{c|}{}
& \textbf{F1} & \textbf{BLEU-1} & \multicolumn{1}{c|}{\textbf{Tokens}}
& \textbf{F1} & \textbf{BLEU-1} & \textbf{Tokens} \\
\hline

Round 1 (w/o R)
& 636 & 45.02 & 37.45 & 2187 & 26.05 & 21.18 & 2509 
& 600 & 43.95 & 36.35 & 2287 & 24.33 & 19.73 & 2534 \\

Round 2 (w/o R)
& 394 & 52.20 & 43.92 & 4353 & 35.82 & 29.85 & 4995 
& 386 & 47.06 & 40.46 & 2864 & 35.50 & 29.34 & 4471 \\

Round 3 (w/o R)
& 140 & 51.53 & 43.78 & 5708 & 40.34 & 33.90 & 6593 
& 152 & 51.88 & 42.83 & 3165 & 38.72 & 32.28 & 5197 \\

Round 4 (w/o R)
& 40 & 37.47 & 32.42 & 5501 & 41.79 & 35.05 & 7322 
& 59 & 44.44 & 36.46 & 3382 & 41.07 & 33.85 & 5638 \\

%Forced Output
%& 320 & 20.04 & 15.48 & 15973 & -- & -- & --
%& 343 & 20.37 & 16.51 & 9243 & -- & -- & -- \\
\hline
Rewriting
& -- & -- & -- & -- & 44.19 & 38.02 & 7618 
& -- & -- & -- & -- & 43.61 & 37.16 & 5799 \\
\hline
\end{tabular}%
}
\end{table*}

\paragraph{Efficiency Analysis.} \label{sec:efficiency}
MGRetrieval achieves a favorable trade-off between effectiveness and efficiency. 
Table~\ref{tab:cost} reports the cost of the main answering LLM. We report the auxiliary keyword extraction cost separately because these operations are consistently handled by the lightweight GPT-4o-mini across all experiments.
During online question answering, MGRetrieval invokes the main LLM 2.23 times per response and consumes 5799 tokens on average, which is substantially lower than the full memory length.
Unlike the baselines MemoryBank, A-Mem, and MemoryOS, MGRetrieval does not reconstruct the memory bank using the main LLM.
Although an auxiliary LLM is used for keyword processing, this process requires only two calls per memory item, consuming 1176.06 tokens on average, and one call per query, consuming 786.43 tokens on average. 
%These operations can already be handled effectively by the lightweight GPT-4o-mini.
Compared with MemoryOS under the full LoCoMo pipeline, {MGRetrieval} achieves a 16.96\% reduction in main LLM token cost and a 14.40\% reduction in total monetary cost with GPT-4o.
In addition, compared with the reflection-based method MemR$^3$, MGRetrieval substantially reduces response time.
Together, these results show that MGRetrieval avoids full memory inputs while maintaining practical token cost and response time.

\subsection{Ablation Study}

\begin{figure}[t]
    \centering
    \includegraphics[width=0.95\columnwidth]{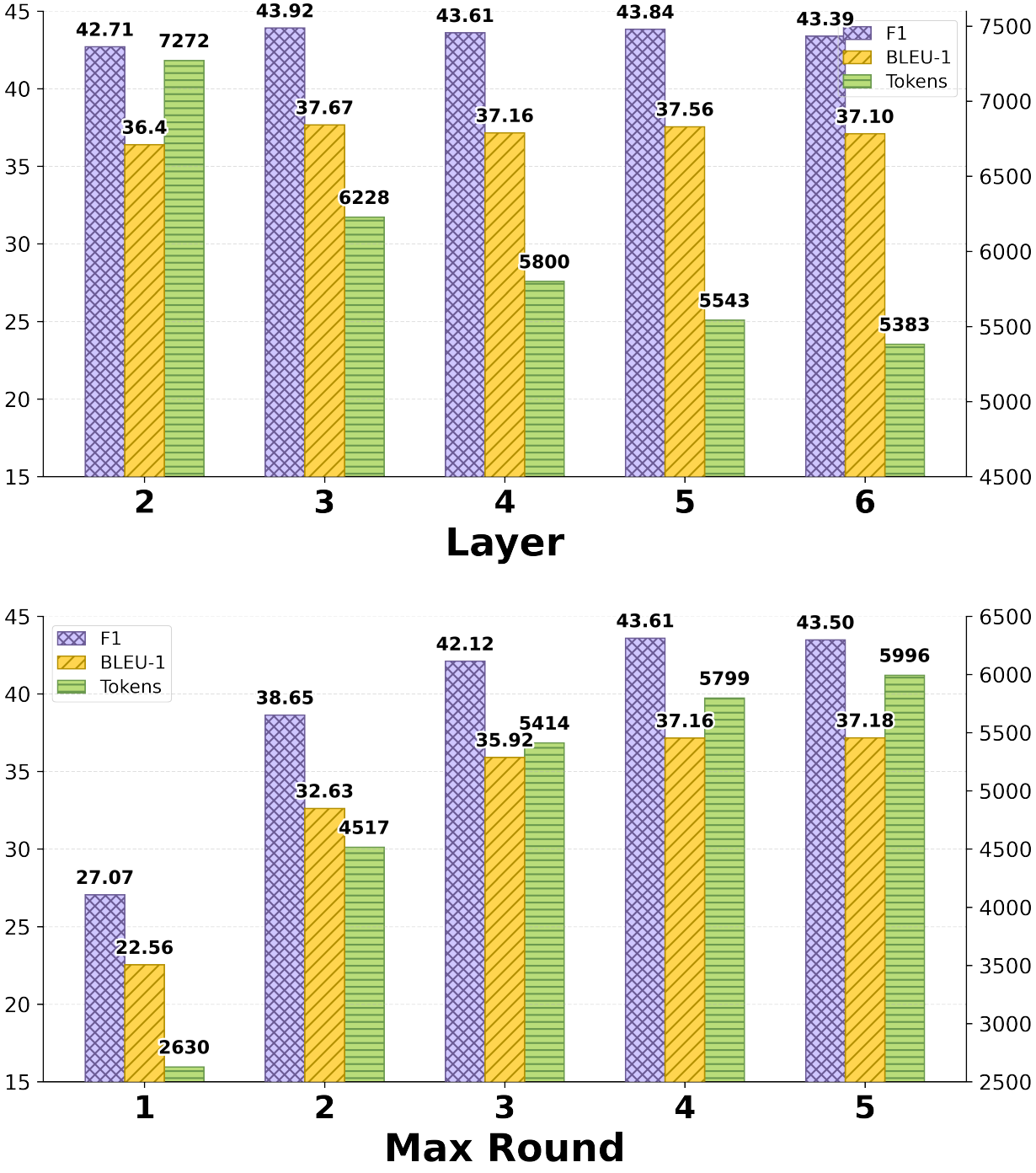}
    \caption{Impact of the pyramid depth $l$ and the maximum number of reflective rounds $r$ with Qwen2.5-14B.}
    \label{fig:parameter}
\end{figure}
\begin{figure*}[t]
    \centering
    \includegraphics[width=\textwidth]{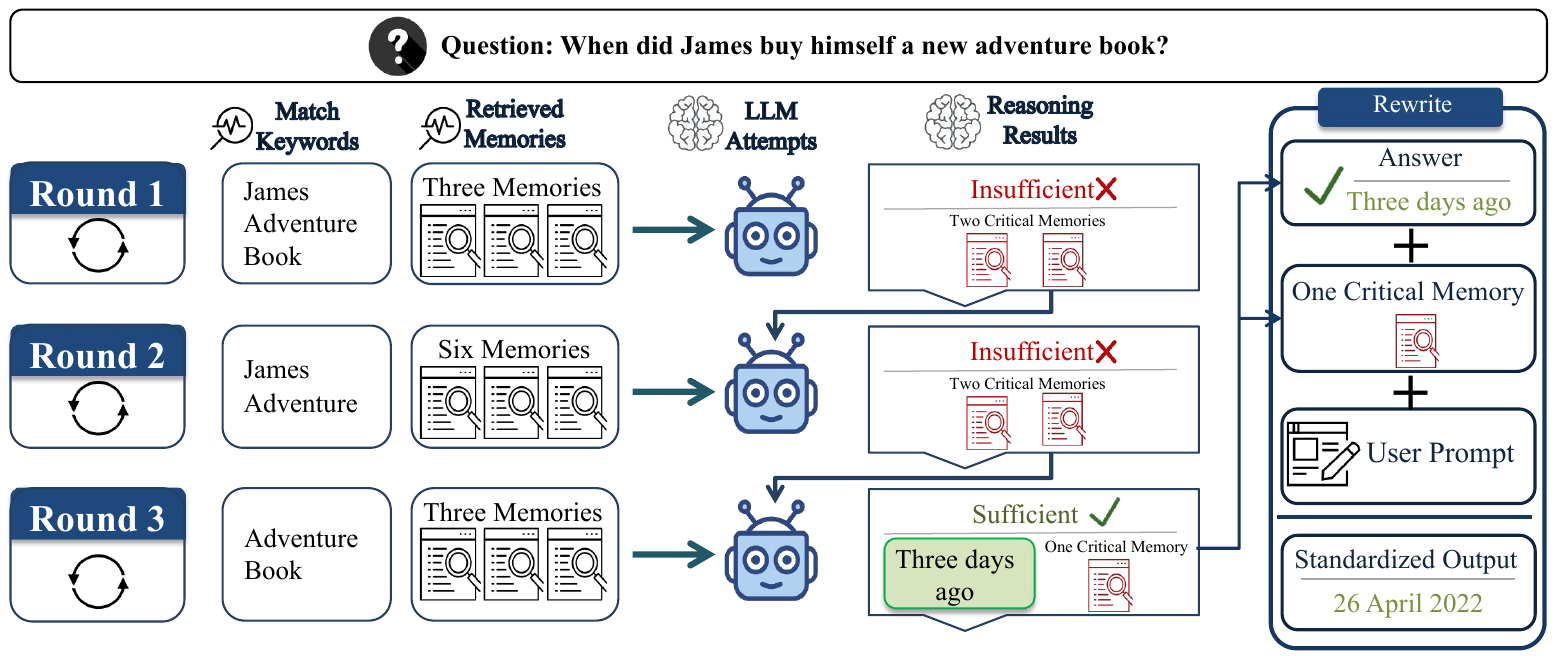}
    \caption{A representative case illustrating the reflection and rewriting processes of MGRetrieval.}
    \label{fig:case}
\end{figure*}
We ablate memory-guided reflective retrieval, redundant memory removal (RMR), and answer rewriting (R), with results reported in Table~\ref{tab:ablation}.

Current performance is computed over samples for which MGRetrieval judges the retrieved memories in the current round to be sufficient for answering the question.
The current scores remain consistently high across rounds.
These consistently high scores indicate that the LLM-assisted sufficiency assessment identifies answerable cases while preserving critical memories.
Full performance is computed over all samples at the current round, regardless of whether the model judges the retrieved memories to be sufficient.
The full scores improve steadily from Round 1 to later rounds and Round 1 also corresponds to the one-shot setting.
This improvement shows that memory-guided retrieval retrieves question-relevant memories missed in earlier rounds.
Together, these results isolate the effect of the memory-guided path on retrieval control.

Removing RMR does not yield clear performance gains, but substantially increases token consumption. This indicates that RMR mainly improves efficiency by filtering repeated memories, without clearly hurting answer quality.

Answer rewriting further improves the final scores.
This improvement shows that rewriting helps normalize answers using only existing critical memories.
Even without rewriting, MGRetrieval retains most of its gains, showing that the main improvement comes from retrieval.
%Notably, rewriting only refines the generated response without introducing new information. 

Overall, the ablation study shows that {MGRetrieval}'s gains mainly come from the memory-guided retrieval mechanism, while RMR and rewriting further improve efficiency and answer normalization. 
Together, these components construct compact inputs that preserve the evidence needed for long-term memory reasoning.

\subsection{Hyperparameter Analysis}

We study two hyperparameters of MGRetrieval: the pyramid depth $l$ and the maximum number of reflective rounds $r$. We vary $l \in \{2,3,4,5,6\}$ and $r \in \{1,2,3,4,5\}$, with results shown in Figure~\ref{fig:parameter}.
Full results can be found in Appendix~\ref{Hyperparameter}.
As $l$ increases, MGRetrieval does not achieve clear performance gains, but token cost is substantially reduced.
Lower pyramid levels cover all the memories retrieved by higher levels, whereas higher levels integrate more keywords, helping MGRetrieval anchor key memories earlier. 
However, excessive keywords may overfit the question, leading to ineffective retrieval at higher levels.
We further observe that, as the maximum number of reflective rounds $r$ increases, MGRetrieval performance first improves and then gradually plateaus. 
This is because only the samples judged as unanswerable are forwarded to the next retrieval round. 
A moderate number of reflective rounds helps improve performance, but as the iterations grow, fewer samples remain to be passed to subsequent rounds.
%We set $l=4$ and $r=4$, which provides the best trade-off between performance and cost.

\subsection{Case}
To intuitively illustrate the role of {MGRetrieval} in constructing concise and sufficient memory contexts through memory-guided retrieval, we present a case study in Figure~\ref{fig:case} to show how it answers a representative question.
{MGRetrieval} first retrieves three memories under the top-level keyword group "James adventure book".
In the first round, the LLM determines that the current memories are insufficient to answer and returns two critical memories. 
It then selects the second keyword group "James adventure" and combines six newly retrieved memories with the critical memories for the second answer.
In the third round, the LLM identifies the relevant memories as sufficient and returns the final answer.
Finally, {MGRetrieval} performs personalized and standardized rewriting based on the answer, the final critical memory, and the user prompt to produce the final output.
In this example, {MGRetrieval} revises the answer "three days ago" into the standardized format "26 April 2022".
Detailed workflow is provided in Appendix~\ref{case}.

\section{Conclusions}

In this work, we propose MGRetrieval, a memory-guided reflective retrieval strategy for long-term conversational LLM agents, to construct concise and sufficient memory contexts.
Instead of relying on one-shot retrieval or reflective retrieval based on LLM-generated queries, MGRetrieval uses the semantic structure of historical memories to guide retrieval paths.
It combines memory-guided retrieval paths with reflective retrieval control, redundant memory removal, and user-defined answer rewriting.
Results on conversational memory benchmarks show that MGRetrieval achieves strong performance with practical token and latency costs.
These findings suggest that the retrieval mechanisms are important in long-term LLM agents and the structure of stored memories can provide useful guidance for retrieval control.

\newpage
\section*{Limitations}
Although MGRetrieval achieves strong results, several limitations remain.

First, MGRetrieval is a retrieval-control strategy rather than a complete memory management system.
It preserves raw memories and uses a lightweight keyword-based index to guide retrieval, but has not yet been integrated with advanced memory organization, updating, and forgetting mechanisms.
Future work may explore comprehensive memory systems built around MGRetrieval.

Second, the LLM-assisted stopping decision does not guarantee that all relevant evidence has been collected. 
This limitation is more obvious for multi-hop, where useful memories may be scattered across different scopes. 
More principled retrieval control could further improve reliability.

Third, the reflective retrieval still relies heavily on the LLM and follows an iterative input process.
As a result, it may require more LLM calls, introducing additional token and response time costs.
Simplifying the reflection mechanism may help alleviate this issue.

Finally, MGRetrieval uses the LLM to adaptively extract and match keywords.
Therefore, the keyword pyramid depends on the quality of keyword extraction and matching.
Noisy, overly generic, or missing keywords may lead to suboptimal retrieval paths, especially when the memory bank contains overlapping topics.
Moreover, excessive keywords may introduce extra LLM calls and semantic ambiguity.
Exploring more accurate semantic matching methods, such as combining keyword matching with vector similarity, may help alleviate this issue.

\section*{Ethical Statement}
This work focuses on developing an efficient retrieval framework for long-term LLM agents. 
All experiments are conducted on public datasets. 
Since MGRetrieval only retrieves from an existing memory bank, it does not collect additional user data.

Nevertheless, long-term agent applications may involve potential ethical risks and personal privacy concerns, especially when memory banks contain sensitive user information.
To mitigate these risks, we recommend adopting strong encryption and privacy-preserving methods during data collection. 
Ethical issues and personal privacy should also be carefully considered when applying MGRetrieval in real-world scenarios. 
Future work should continue to prioritize ethical considerations.

\section*{AI Assistance Statement}
AI tools were used only for grammar correction and spell checking. All research ideas, methods, theoretical claims, code, experiments, analyses, and citations were produced and verified by the authors.
\section*{Acknowledgments}

\bibliography{custom}
\appendix
\section{Experiment}
\subsection{Detailed Dataset Descriptions} \label{Datasets}
We use the latest standards from the codebases provided with these datasets and consistently follow their defined evaluation procedures.
\paragraph{LoCoMo.}
LoCoMo \cite{maharana2024evaluating} is a dataset of very long-term conversations, with each conversation encompassing 600 turns on average, over up to 32 sessions. 
The released LoCoMo benchmark consists of ten conversations, each annotated for the question-answering and event-summarization tasks.
In this work, we focus on four reasoning categories in the LoCoMo question-answering task to evaluate {MGRetrieval} from multiple perspectives: 
(1) single-hop questions based on a single session; (2) multi-hop questions synthesizing information across sessions; (3) temporal questions requiring time-aware reasoning; and (4) open-domain knowledge questions combining dialogue context with commonsense or world knowledge.
\paragraph{GVD.}
The GVD~\cite{zhong2024memorybank} evaluation data consist of a memory bank built from 10-day conversations. 
These conversations involve 15 virtual users with diverse personalities. 
ChatGPT is used to simulate the users and generate long-term dialogue contexts across various topics.

\subsection{Evaluation Metrics} \label{Metrics}
For the LoCoMo dataset, we use F1 and BLEU-1 as the primary metric. In addition, we report ROUGE-L, ROUGE-2, and METEOR to provide complementary evaluation of generation quality. For the GVD evaluation data derived from MemoryBank, we follow MemoryBank’s metric definitions, including memory retrieval accuracy, response correctness, and contextual coherence. We use DeepSeek-R1 as an automatic judge to assign these scores.
\paragraph{F1.}
F1 is the harmonic mean of precision and recall. It measures token overlap between the predicted answer and the reference answer.
This metric is particularly valuable for balancing completeness and accuracy in responses:
\begin{equation}
F1 = \frac{2PR}{P + R},
\end{equation}
where
\begin{equation}
P = \frac{| \hat{Y} \cap Y |}{|\hat{Y}|}, \qquad
R = \frac{| \hat{Y} \cap Y |}{|Y|}.
\end{equation}
Here, $\hat{Y}$ and $Y$ denote the token set of the predicted and reference answers. If multiple references are available, we take the maximum F1 over all references and then average over all questions:
\begin{equation}
F1_{\mathrm{avg}} = \frac{1}{N}\sum_{i=1}^{N} \max_{j} F1(\hat{Y}_i, Y_{ij}).
\end{equation}

\paragraph{BLEU-1.}
BLEU-1 measures unigram overlap between a generated answer and a reference answer. It is less strict than exact matching and is widely used for generative evaluation.
\begin{equation}
\mathrm{BLEU\mbox{-}1} = BP \cdot p_1,
\end{equation}
where $p_1$ denotes the modified unigram precision:
\begin{equation}
p_1 =
\frac{
\sum_{w}
\min\!\bigl(
C_{\mathrm{cand}}(w),
C_{\mathrm{ref}}^{\max}(w)
\bigr)
}{
\sum_{w}
C_{\mathrm{cand}}(w)
}.
\end{equation}
Here, $C_{\mathrm{cand}}(w)$ is the count of unigram $w$ in the candidate answer, and $C_{\mathrm{ref}}^{\max}(w)$ is the maximum count of $w$ in the reference answers. 
\begin{equation}
BP =
\begin{cases}
1, & c > r, \\
e^{1-r/c}, & c \le r,
\end{cases}
\end{equation}
where $c$ and $r$ denote the candidate length and effective reference length.

\paragraph{ROUGE-L.}
ROUGE-L~\cite{lin2004rouge} measures the longest common subsequence (LCS) between the generated answer and the reference answer:
\begin{equation}
\mathrm{ROUGE\mbox{-}L} = \frac{(1+\beta^2) R_l P_l}{R_l + \beta^2 P_l},
\end{equation}
where
\begin{equation}
R_l = \frac{\mathrm{LCS}(\hat{Y}, Y)}{|Y|}, \qquad
P_l = \frac{\mathrm{LCS}(\hat{Y}, Y)}{|\hat{Y}|}.
\end{equation}
Here, $\hat{Y}$ and $Y$ denote the generated and reference answers, respectively, and $\mathrm{LCS}(\hat{Y}, Y)$ denotes the length of their longest common subsequence.

\paragraph{ROUGE-2.}
ROUGE-2~\cite{lin2004rouge} measures bigram overlap between the generated answer and the reference answer:
\begin{equation}
\mathrm{ROUGE\mbox{-}2}
=
\frac{
\sum_{b \in Y}
\min\!\bigl(c_r(b),\, c_c(b)\bigr)
}{
\sum_{b \in Y}
c_r(b)
}.
\end{equation}
Here, $c_r(b)$ and $c_c(b)$ denote the counts of bigram $b$ in the reference answer $Y$ and the generated answer $\hat{Y}$, respectively.

\paragraph{METEOR.}
METEOR~\cite{banerjee2005meteor} measures unigram alignment between the generated answer and the reference answer while taking synonyms and paraphrases into account:
\begin{equation}
\mathrm{METEOR} = F_{\mathrm{mean}} \cdot (1 - \mathrm{Penalty}),
\end{equation}
where
\begin{equation}
F_{\mathrm{mean}} = \frac{10PR}{R + 9P},
\end{equation}
and
\begin{equation}
\mathrm{Penalty} = 0.5 \cdot \left(\frac{ch}{m}\right)^3.
\end{equation}
Here, $P$ and $R$ denote unigram precision and recall, $ch$ is the number of chunks, and $m$ is the number of matched unigrams.

\paragraph{Tokens and Calls.}
We divide the full LoCoMo pipeline into memory bank construction and question answering. 
For each stage, we record the number of LLM calls and estimate the token consumption, averaged per memory item and per question, respectively. 
Let w and s denote the numbers of words and symbols, the token count is estimated as:
\begin{equation}
tokens = 1.1w + 0.35s.
\end{equation}

\paragraph{Time (s).}
For each evaluation question, response time (s) is measured as the elapsed time from sending the first request to receiving the final response.

\paragraph{Memory Retrieval Accuracy (Acc.).}
Acc. measures whether the system retrieves the memory required to answer a query. In GVD-style evaluation, it is annotated with binary labels $r_i \in \{0,1\}$:
\begin{equation}
Acc. = \frac{1}{N}\sum_{i=1}^{N} r_i.
\end{equation}

\paragraph{Response Correctness (Corr.).}
Corr. measures whether the generated response correctly answers the question at the semantic level. It is annotated on a three-level scale $c_i \in \{0, 0.5, 1\}$:
\begin{equation}
Corr. = \frac{1}{N}\sum_{i=1}^{N} c_i.
\end{equation}

\paragraph{Contextual Coherence (Cohe.).}
Cohe. measures whether the generated response is coherent with the dialogue context. It is also annotated on a three-level scale $h_i \in \{0, 0.5, 1\}$:
\begin{equation}
Cohe. = \frac{1}{N}\sum_{i=1}^{N} h_i.
\end{equation}

\subsection{Detailed Baseline Descriptions} \label{Baseline}
\paragraph{FULL.}
FULL does not employ any memory management mechanism, including memory storage, deletion, updating, and retrieval. 
It directly uses the full historical memories.

\paragraph{RAG.}
RAG \cite{karpukhin2020dense} does not employ any memory storage, deletion, or updating mechanism. 
It only uses the bce-embedding-base\_v1 model to retrieve the top 20 most relevant memories based on vector similarity for reasoning.

\paragraph{MemR$^3$.}
We adopt the reflection-based RAG method in MemR$^3$ \cite{du2025memr}. It transforms retrieval into a closed-loop process of Retrieve–Reflect–Answer, allowing the model to adaptively retrieve memories and produce answers.

\paragraph{MemoryBank.}
MemoryBank \cite{zhong2024memorybank} maintains long-term memory by preserving historical memories, event-level summaries, and user profile information, while adjusting memory strength with a forgetting mechanism inspired by the Ebbinghaus curve.
It encodes memory units into dense representations and selects the most relevant memories according to the query context.

\paragraph{A-Mem.}
A-Mem \cite{xu2025mem} organizes each memory as a structured note, which is represented by contextual descriptions, keywords, and tags.
Existing notes evolve as interactions proceed.
The agent relies on dynamic indexing and semantic linking among related notes to retrieve relevant memories through an evolving network.

\paragraph{MemoryOS.}
MemoryOS \cite{kang2025memory} adopts a hierarchical memory architecture that separates information into short-term, mid-term, and long-term personal memory. It updates these layers through cross-level transfer mechanisms.
For retrieval, it gathers semantically relevant context from this hierarchical memory structure.

\subsection{Implementation Details} \label{Implementation Details}
We conduct all experiments with Python 3.10 and PyTorch 2.4.1+cu121 on an NVIDIA RTX 6000 GPU with 24GB. 
For LLMs, we use the official APIs and set the temperature to 0.0.

For all baselines, we use the latest publicly available versions and follow the experimental configurations reported in their original papers. 
For settings that are not specified in the papers, we follow the default configurations provided in their code repositories.

For MGRetrieval, we set the maximum number of retrieval iterations to 4 in all experiments. 
The maximum number of query keywords selected by the prompt should be aligned with the pyramid depth.
For experiments with pyramid depth $l$, the query keyword selection prompt is configured to select up to $l$ keywords.
For main experiments, we set the pyramid depth to 4.
We consistently use GPT-4o-mini for keyword extraction and keyword matching, while the main LLM is used only for response generation. 
Therefore, the token consumption of keyword extraction and matching is not included in the reported main LLM token cost.
All prompts can be found in Appendix~\ref{promp}.
The rewriting module only normalizes the final answer using the existing critical memories and a few formatting examples, without retrieving additional evidence.
Therefore, it does not compromise the fairness of the evaluation.

\subsection{Empirical Results} \label{Results}

\begin{table*}[tb!]
\centering
\caption{
Experimental results of Qwen2.5-14B and Qwen3-14B on the LoCoMo dataset across four categories.
}
\label{tab:qwen}

\setlength{\tabcolsep}{2.5pt}
\renewcommand{\arraystretch}{0.95}
\resizebox{\textwidth}{!}{%
\begin{tabular}{ccl|ccc|ccc|ccc|ccc}
\hline
\multicolumn{2}{c}{\multirow{3}{*}{\textbf{Model}}} 
& \multicolumn{1}{c|}{\multirow{3}{*}{\textbf{Method}}} 
& \multicolumn{12}{c}{\textbf{Category}} \\ 
\cline{4-15}
\multicolumn{2}{c}{} 
& \multicolumn{1}{c|}{} 
& \multicolumn{3}{c|}{\textbf{Multi Hop}} 
& \multicolumn{3}{c|}{\textbf{Temporal}} 
& \multicolumn{3}{c|}{\textbf{Open Domain}} 
& \multicolumn{3}{c}{\textbf{Single Hop}} \\
\multicolumn{2}{c}{} 
& \multicolumn{1}{c|}{} 
& \textbf{ROUGE-L} & \textbf{ROUGE-2} & \multicolumn{1}{c|}{\textbf{METEOR}} 
& \textbf{ROUGE-L} & \textbf{ROUGE-2} & \multicolumn{1}{c|}{\textbf{METEOR}} 
& \textbf{ROUGE-L} & \textbf{ROUGE-2} & \multicolumn{1}{c|}{\textbf{METEOR}} 
& \textbf{ROUGE-L} & \textbf{ROUGE-2} & \textbf{METEOR} \\ 
\hline

\multirow{14}{*}{\textbf{\rotatebox{90}{Qwen}}} 
& \multicolumn{1}{c|}{\multirow{7}{*}{\textbf{\rotatebox{90}{2.5-14B}}}} 
& \textsc{MemoryBank} 
& 19.47 & 8.63 & \multicolumn{1}{c|}{18.47} 
& 10.22 & 4.11 & \multicolumn{1}{c|}{7.60} 
& 4.36 & 1.28 & \multicolumn{1}{c|}{3.41} 
& 38.52 & 34.98 & 42.88 \\

& \multicolumn{1}{c|}{} 
& \textsc{FULL} 
& 21.14 & 8.55 & \multicolumn{1}{c|}{16.75} 
& 18.36 & 5.99 & \multicolumn{1}{c|}{13.27} 
& 15.36 & 3.16 & \multicolumn{1}{c|}{9.83} 
& 42.50 & 31.29 & 40.46 \\

& \multicolumn{1}{c|}{} 
& \textsc{RAG} 
& 17.92 & 6.72 & \multicolumn{1}{c|}{13.62} 
& 25.65 & 7.95 & \multicolumn{1}{c|}{18.14} 
& 13.82 & 3.18 & \multicolumn{1}{c|}{9.35} 
& 41.21 & 28.89 & 37.44 \\

& \multicolumn{1}{c|}{} 
& \textsc{A-Mem} 
& 29.90 & 12.29 & \multicolumn{1}{c|}{17.44} 
& 24.33 & 7.06 & \multicolumn{1}{c|}{14.11} 
& 14.31 & 2.37 & \multicolumn{1}{c|}{6.64} 
& 40.29 & 25.76 & 33.60 \\

& \multicolumn{1}{c|}{} 
& \textsc{MemR$^3$} 
& 28.19 & 12.61 & \multicolumn{1}{c|}{\textbf{25.05}} 
& 12.20 & 4.08 & \multicolumn{1}{c|}{9.25} 
& 19.13 & 6.58 & \multicolumn{1}{c|}{14.35} 
& 50.60 & \textbf{37.91} & \textbf{46.77} \\

& \multicolumn{1}{c|}{} 
& \textsc{MemoryOS} 
& 31.20 & 12.24 & \multicolumn{1}{c|}{23.58} 
& 29.05 & 10.29 & \multicolumn{1}{c|}{20.19} 
& 23.43 & \textbf{8.13} & \multicolumn{1}{c|}{16.05} 
& 43.89 & 28.62 & 37.22 \\

& \multicolumn{1}{c|}{} 
& \textsc{MGRetrieval} 
& \textbf{31.29} & \textbf{12.67} & \multicolumn{1}{c|}{22.05} 
& \textbf{34.77} & \textbf{17.20} & \multicolumn{1}{c|}{\textbf{30.68}} 
& \textbf{27.18} & 6.94 & \multicolumn{1}{c|}{\textbf{16.32}} 
& \textbf{51.26} & 35.16 & 46.20 \\ 
\cline{2-15}

& \multicolumn{1}{c|}{\multirow{7}{*}{\textbf{\rotatebox{90}{3-14B}}}} 
& \textsc{MemoryBank} 
& 30.29 & 10.93 & \multicolumn{1}{c|}{25.41} 
& 21.34 & 11.84 & \multicolumn{1}{c|}{18.56} 
& 11.83 & 2.34 & \multicolumn{1}{c|}{8.76} 
& 42.07 & 31.50 & 39.51 \\

& \multicolumn{1}{c|}{} 
& \textsc{FULL} 
& 22.15 & 8.61 & \multicolumn{1}{c|}{17.57} 
& 33.72 & 9.17 & \multicolumn{1}{c|}{21.89} 
& 15.16 & 3.21 & \multicolumn{1}{c|}{9.57} 
& 43.60 & 29.77 & 39.12 \\

& \multicolumn{1}{c|}{} 
& \textsc{RAG} 
& 20.02 & 7.07 & \multicolumn{1}{c|}{15.40} 
& \textbf{36.02} & 9.71 & \multicolumn{1}{c|}{22.95} 
& 11.23 & 4.14 & \multicolumn{1}{c|}{7.72} 
& 45.21 & 29.25 & 38.86 \\

& \multicolumn{1}{c|}{} 
& \textsc{A-Mem} 
& 28.47 & 11.70 & \multicolumn{1}{c|}{14.13} 
& 23.69 & 7.63 & \multicolumn{1}{c|}{6.70} 
& 12.77 & 2.92 & \multicolumn{1}{c|}{5.73} 
& 42.99 & 27.65 & 31.05 \\

& \multicolumn{1}{c|}{} 
& \textsc{MemR$^3$} 
& 26.09 & 10.12 & \multicolumn{1}{c|}{20.78} 
& 20.63 & 7.45 & \multicolumn{1}{c|}{14.61} 
& 13.96 & 5.41 & \multicolumn{1}{c|}{9.80} 
& 44.71 & 31.29 & 39.46 \\

& \multicolumn{1}{c|}{} 
& \textsc{MemoryOS} 
& \textbf{31.07} & \textbf{14.87} & \multicolumn{1}{c|}{\textbf{26.22}} 
& 30.95 & 11.23 & \multicolumn{1}{c|}{21.93} 
& 18.00 & 4.73 & \multicolumn{1}{c|}{11.65} 
& 43.43 & 29.87 & 38.31 \\

& \multicolumn{1}{c|}{} 
& \textsc{MGRetrieval} 
& 26.67 & 11.04 & \multicolumn{1}{c|}{20.62} 
& 35.06 & \textbf{14.09} & \multicolumn{1}{c|}{\textbf{28.22}} 
& \textbf{23.03} & \textbf{5.67} & \multicolumn{1}{c|}{\textbf{13.95}} 
& \textbf{49.36} & \textbf{32.20} & \textbf{43.80} \\ 
\hline

\end{tabular}%
}
\end{table*}
We further provide a detailed comparison of the five evaluation metrics on the LoCoMo dataset using Qwen2.5-14B and Qwen3-14B, as shown in Table~\ref{tab:qwen}.

Our evaluation shows that MGRetrieval achieves competitive performance across five metrics while maintaining computational and cost efficiency. 
Compared with memory management methods, including MemoryBank, A-Mem, and MemoryOS, MGRetrieval obtains higher average scores in most settings. 
Compared with retrieval-centric strategies such as RAG and MemR3, it improves the trade-off between effectiveness and efficiency, especially by reducing the response time of reflective retrieval.

For example, with Qwen2.5-14B, MGRetrieval reduces token consumption by 86.63\% relative to directly using the full context as input, while achieving relative gains of 17.74\% in F1 and 21.64\% in BLEU-1. 
Compared with the comprehensive memory management system MemoryOS, it further reduces token consumption by 16.96\%, together with relative improvements of 13.13\% in F1 and 12.30\% in BLEU-1. 
Relative to the advanced retrieval-centric strategy MemR$^3$, MGRetrieval improves F1 by 19.97\% and BLEU-1 by 20.45\%. 
More importantly, compared with MemR$^3$, which also incorporates a reflective mechanism, it markedly reduces response time from 12.77 s to 6.98 s.
These results indicate that MGRetrieval improves the effectiveness and efficiency trade-off by guiding retrieval with a lightweight keyword structure.
Experimental results also show the effectiveness of MGRetrieval in constructing concise yet sufficient input for LLM reasoning.

This underscores the importance of an effective retrieval strategy in memory management systems for LLM agents and also suggests the strong potential of MGRetrieval for LLM agents in long-term real-world applications.

\subsection{Hyperparameter Analysis} \label{Hyperparameter}
\begin{figure}[t]
    \centering
    \includegraphics[width=0.95\columnwidth]{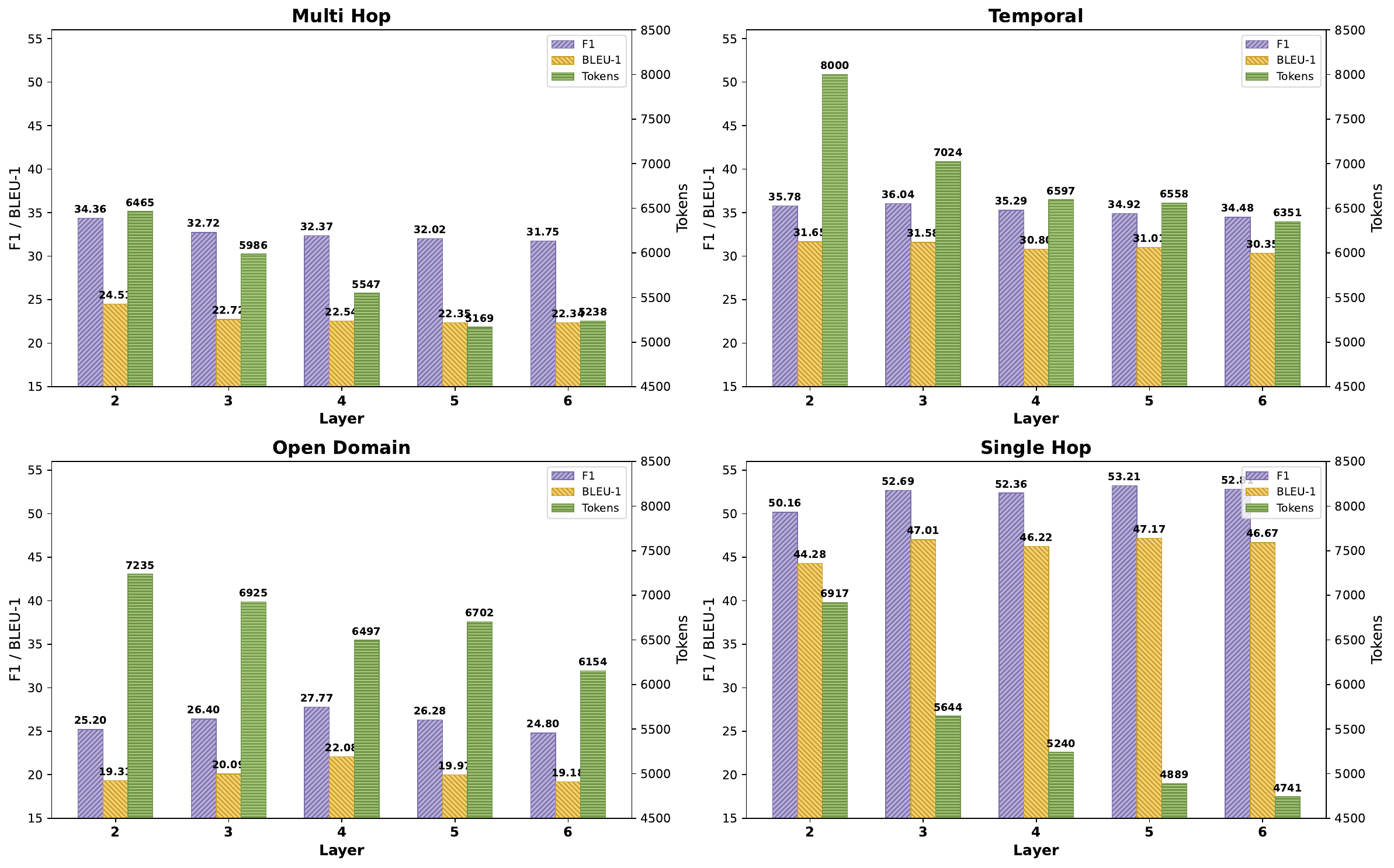}
    \caption{Impact of the pyramid depth $l$ across different task categories with Qwen2.5-14B as the base model.}
    \label{fig:layer}
\end{figure}
\begin{figure}[t]
    \centering
    \includegraphics[width=0.95\columnwidth]{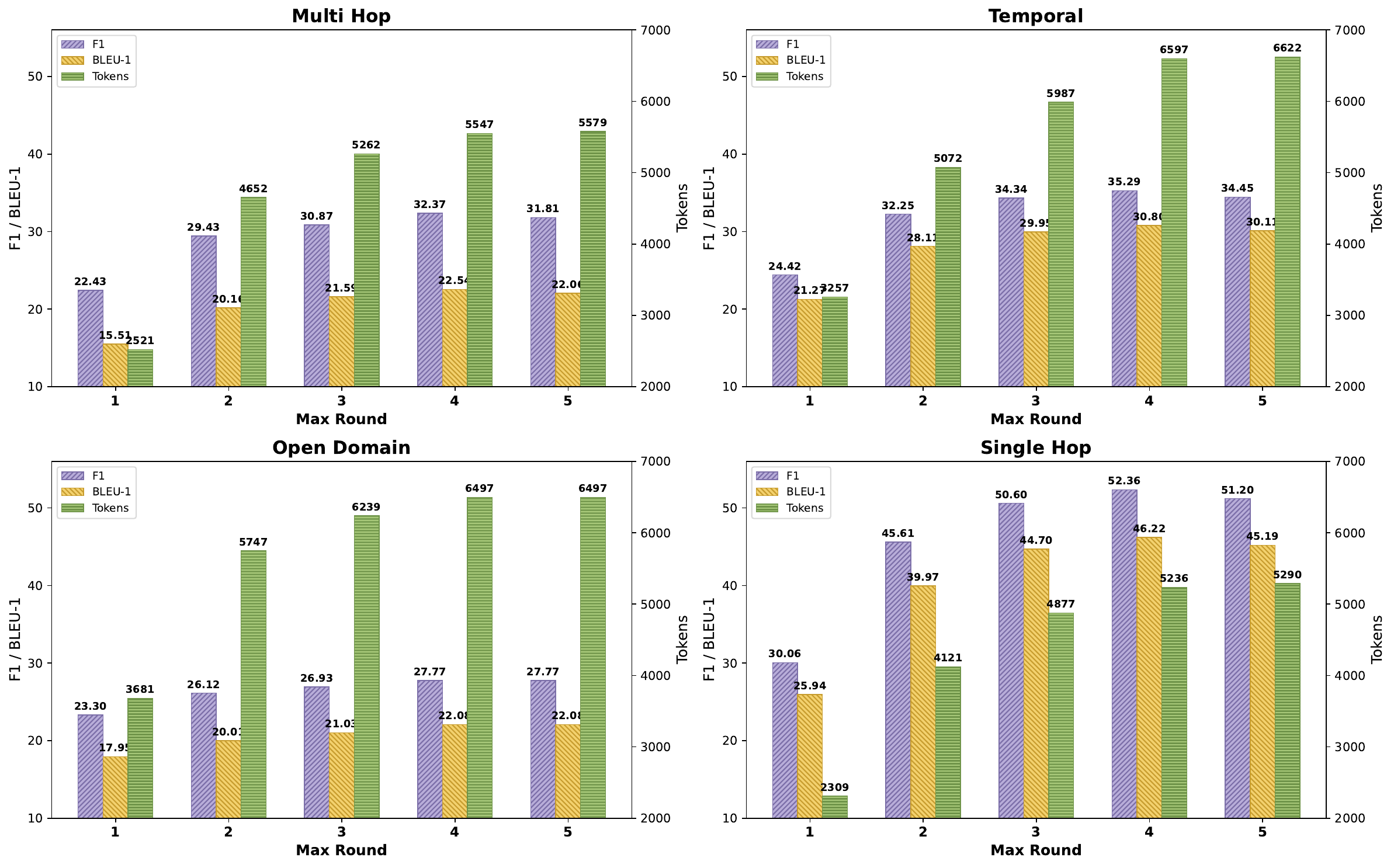}
    \caption{Impact of the maximum number of reflective rounds $r$ across different task categories with Qwen2.5-14B as the base model.}
    \label{fig:maxround}
\end{figure}
MGRetrieval has two key hyperparameters: the maximum number of reflective rounds r and the pyramid depth l. 
We analyze their effects on model performance in Figures \ref{fig:layer} and \ref{fig:maxround}. 
As the number of reflective rounds increases, both performance and token cost rise, but the growth gradually slows down. 
This is because each round filters out samples whose memories are already sufficient, leaving fewer cases to be corrected by newly retrieved memories. 
Thus, early rounds bring clear performance gains with higher token cost, while later rounds gradually approach convergence. Increasing the pyramid depth generally reduces token cost, while performance varies only slightly across task categories. 
Higher pyramid levels combine more keywords and retrieve more specific memory groups, helping MGRetrieval anchor key memories earlier and avoid large memory inputs. 
However, excessively high pyramid levels may overfit the current question and lead to diminishing returns. 
We set r=4 and l=4, which provides a balanced trade-off between token cost and performance.
\section{Case} \label{case}
\begin{figure*}[t]
    \centering
    \includegraphics[width=\textwidth]{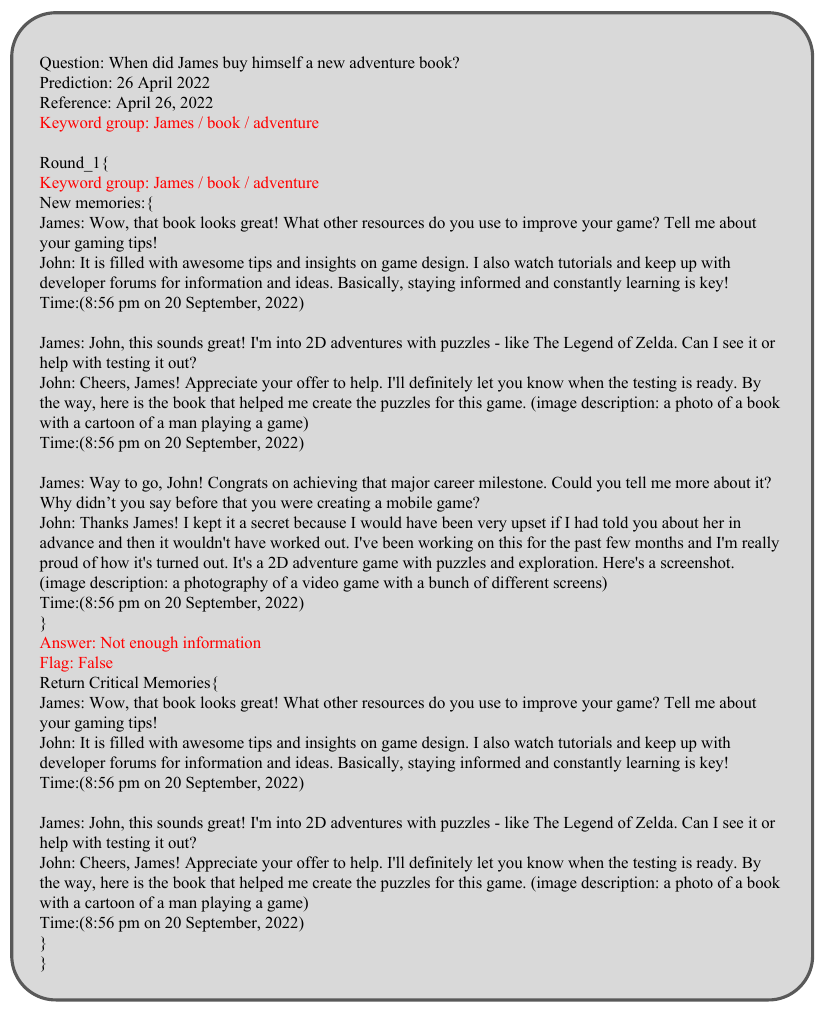}
    \caption{A case study illustrating how {MGRetrieval} answers a representative question.}
    \label{fig:case1}
\end{figure*}

\begin{figure*}[t]
    \centering
    \includegraphics[width=\textwidth]{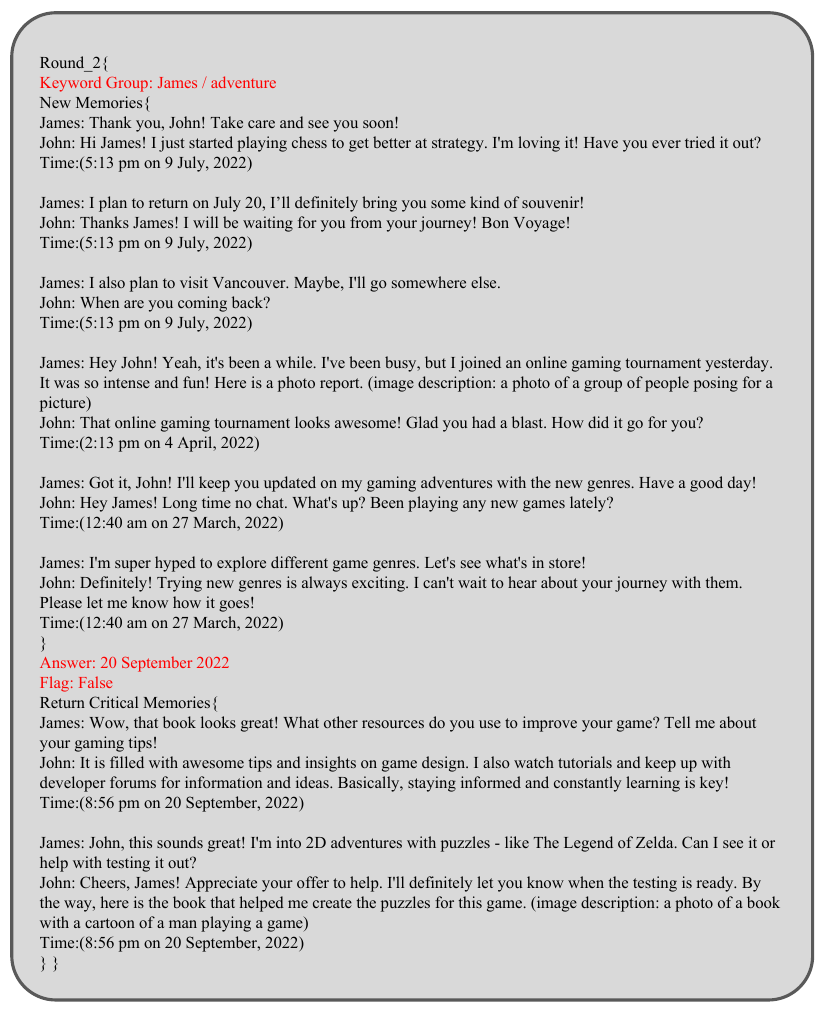}
    \caption{A case study illustrating how {MGRetrieval} answers a representative question.}
    \label{fig:case2}
\end{figure*}

\begin{figure*}[t]
    \centering
    \includegraphics[width=\textwidth]{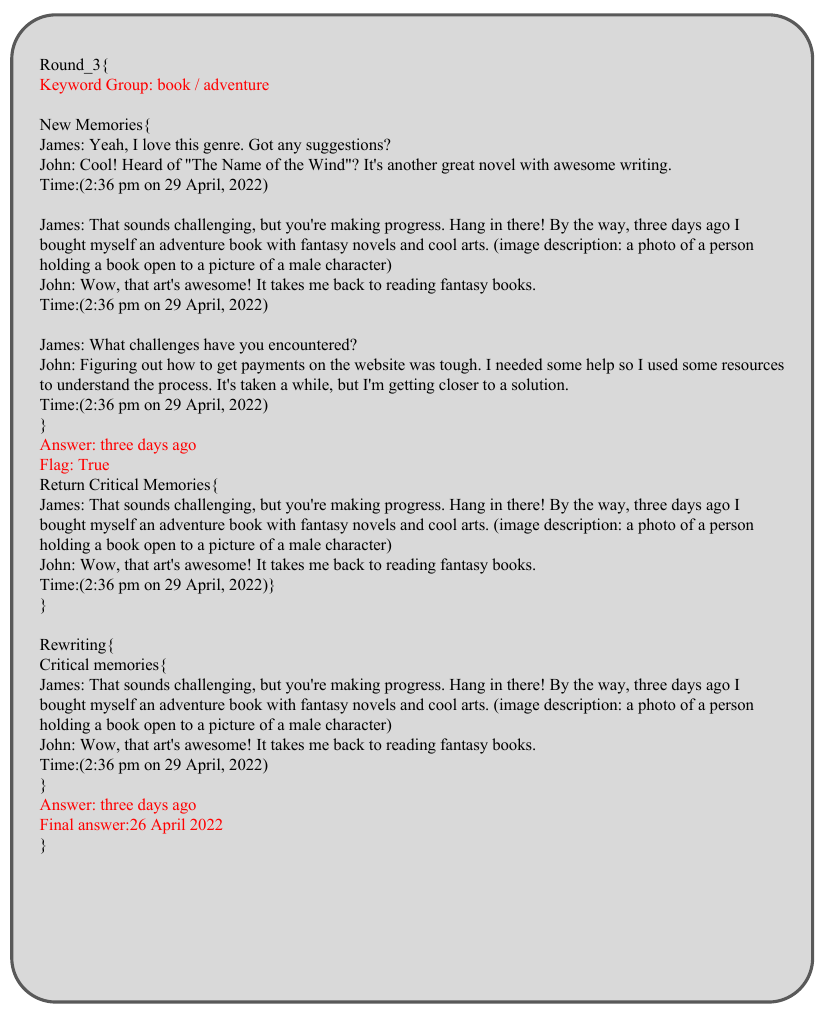}
    \caption{A case study illustrating how {MGRetrieval} answers a representative question.}
    \label{fig:case3}
\end{figure*}

We present a case study to show how {MGRetrieval} answers a representative question. 
The full workflow is presented in Figures \ref{fig:case1}, \ref{fig:case2}, and \ref{fig:case3}.
This case highlights the crucial role of the reflective strategy in retrieving effective memories.
\section{Prompt} \label{promp}
\begin{figure*}[t]
    \centering
    \includegraphics[width=\textwidth]{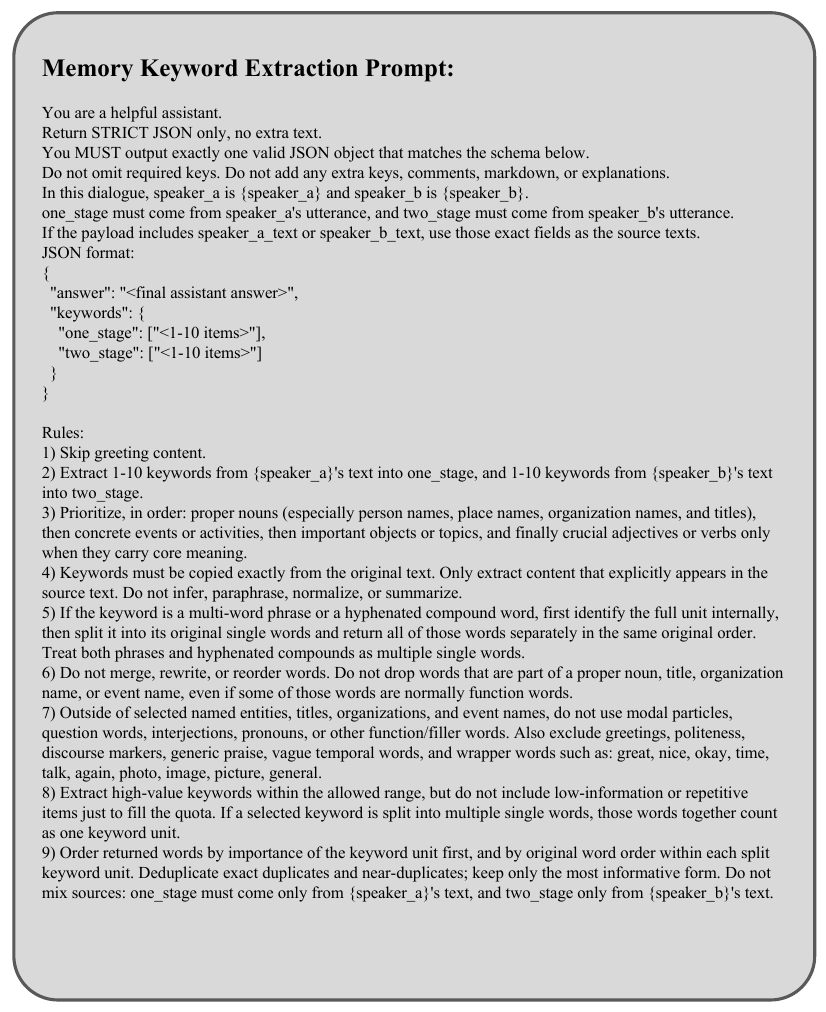}
    \caption{The prompt for memory keyword extraction.}
    \label{fig:prompt1}
\end{figure*}

\begin{figure*}[t]
    \centering
    \includegraphics[width=\textwidth]{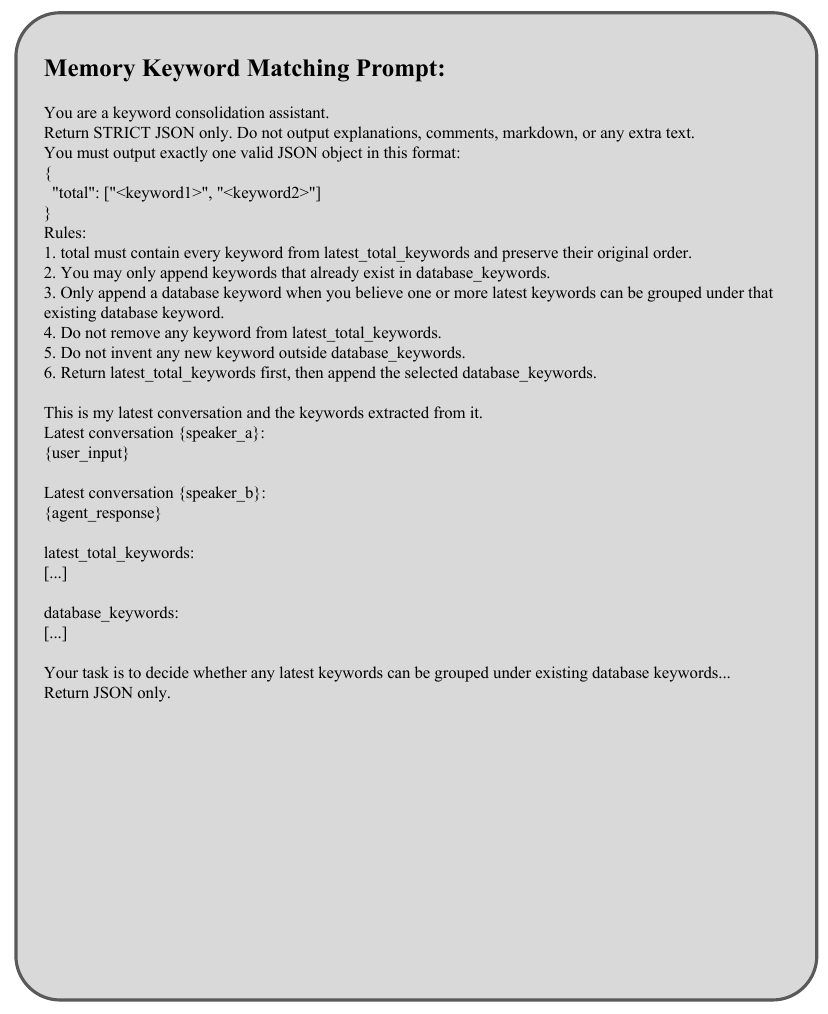}
    \caption{The prompt for memory keyword matching.}
    \label{fig:prompt2}
\end{figure*}

\begin{figure*}[t]
    \centering
    \includegraphics[width=\textwidth]{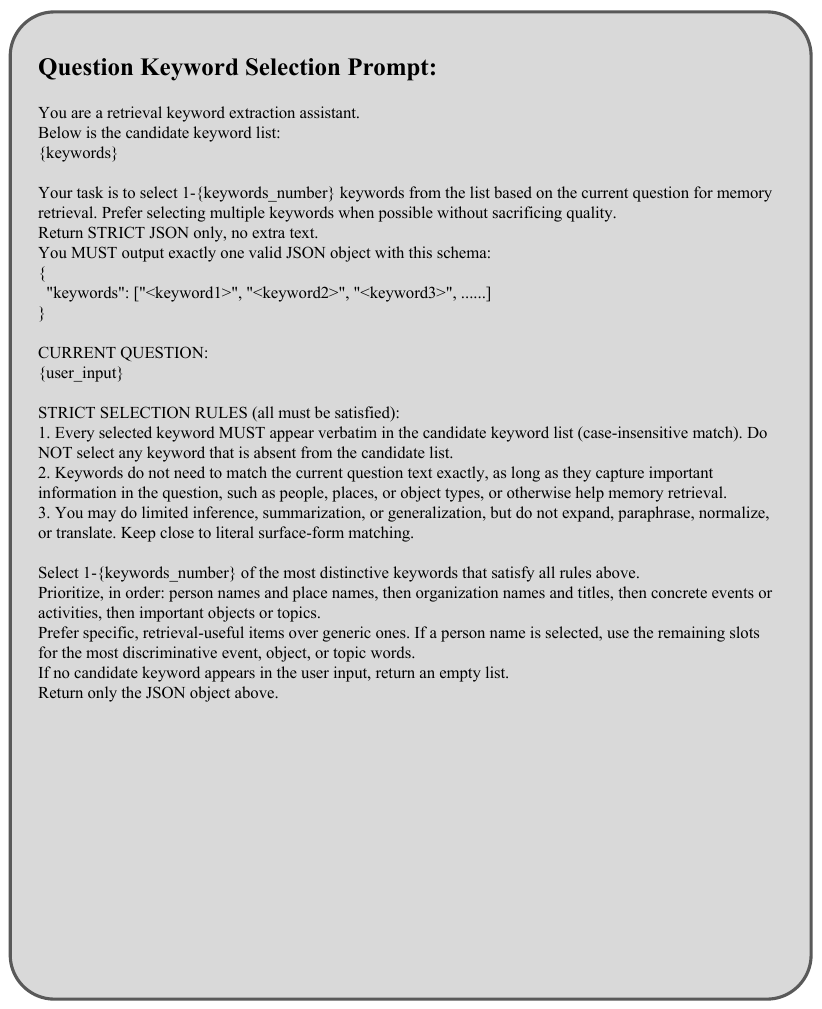}
    \caption{The prompt for question keyword selection.}
    \label{fig:prompt3}
\end{figure*}

\begin{figure*}[t]
    \centering
    \includegraphics[width=\textwidth]{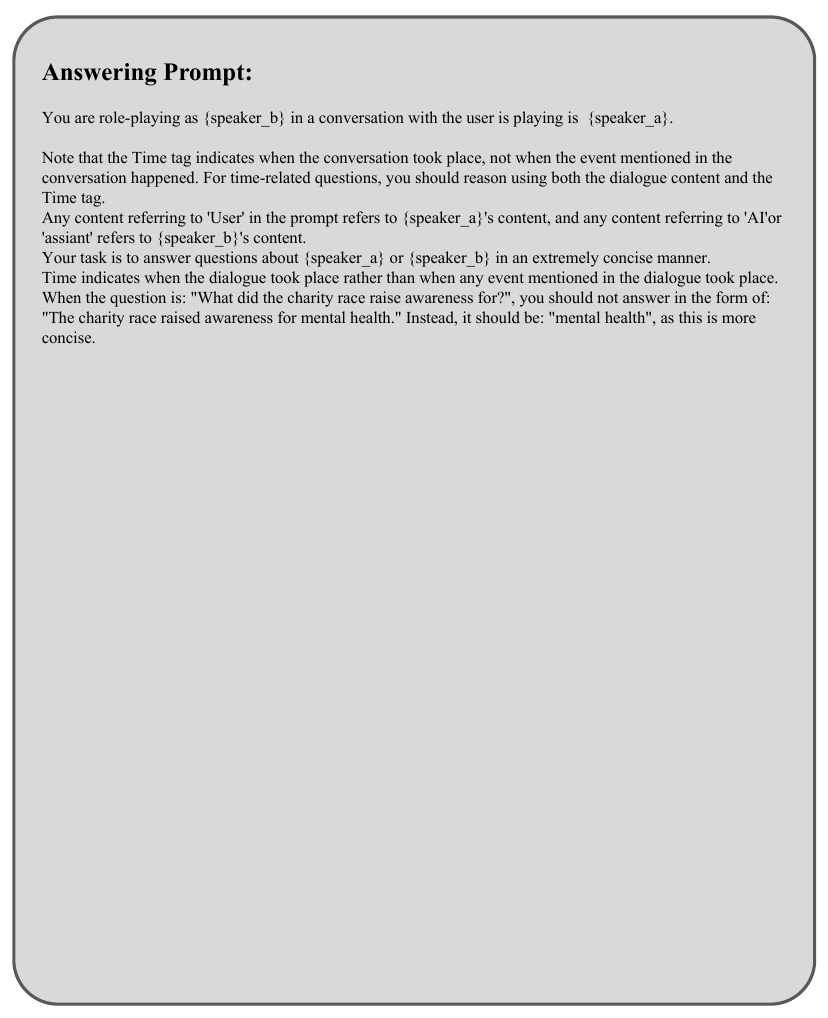}
    \caption{The system prompt for answering.}
    \label{fig:prompt4}
\end{figure*}

\begin{figure*}[t]
    \centering
    \includegraphics[width=\textwidth]{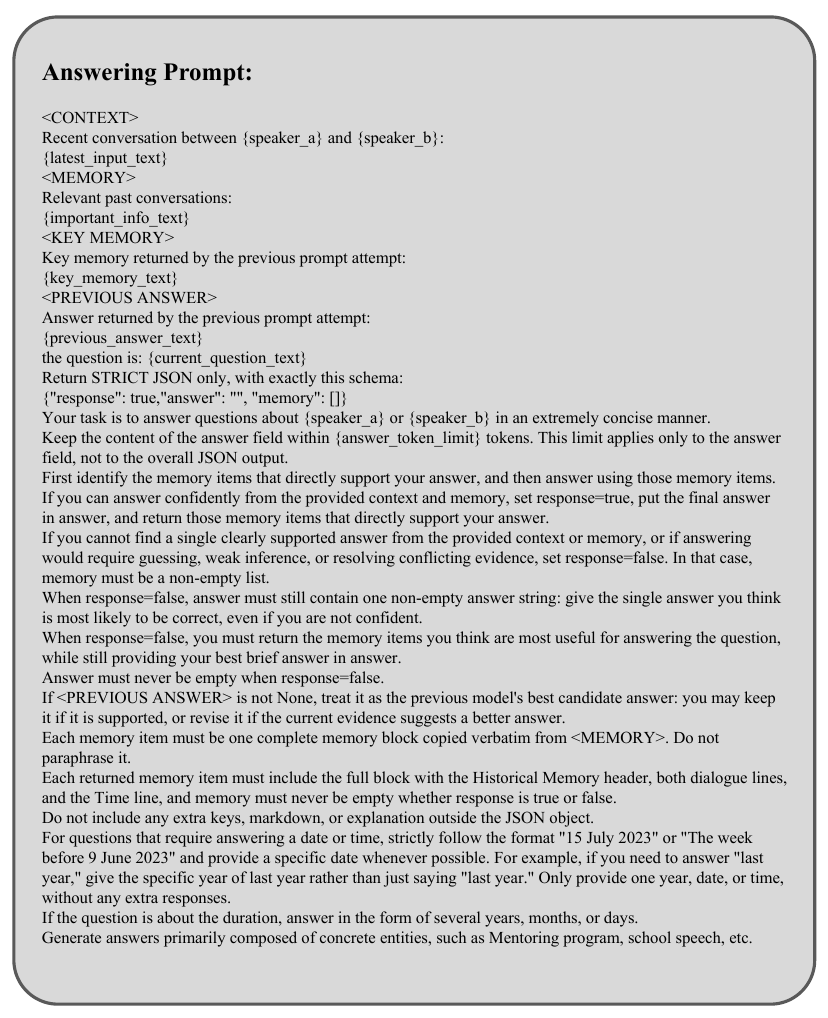}
    \caption{The user prompt for answering.}
    \label{fig:prompt5}
\end{figure*}

\begin{figure*}[t]
    \centering
    \includegraphics[width=\textwidth]{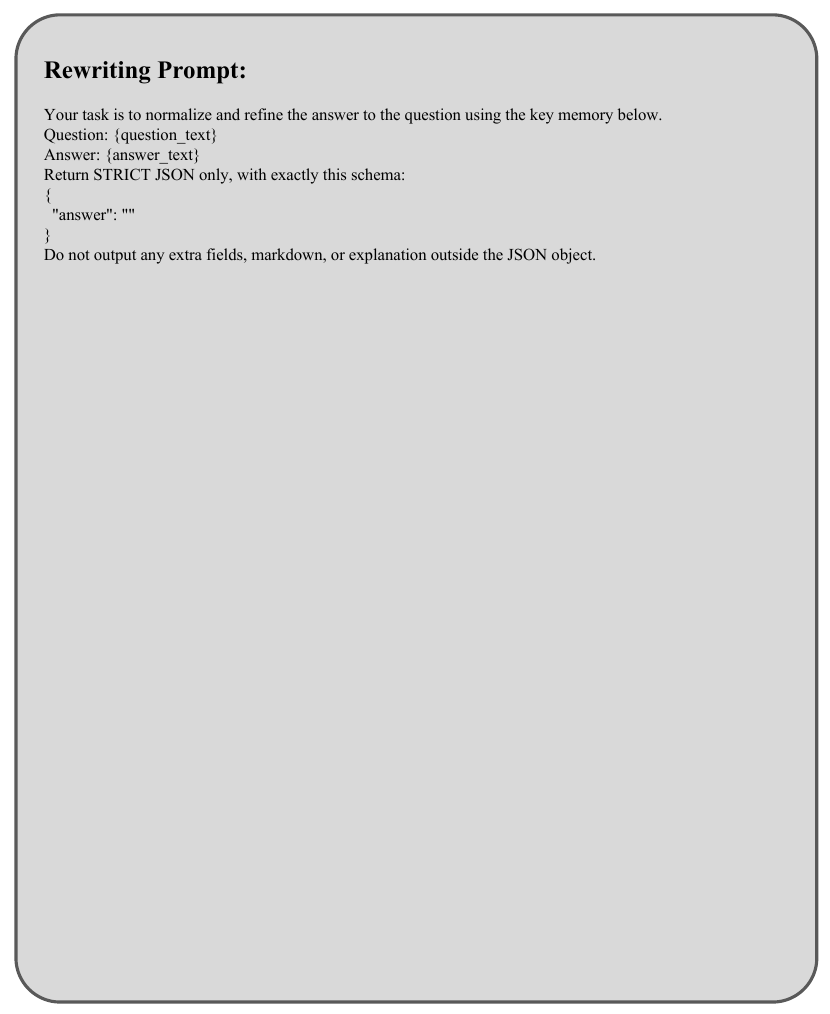}
    \caption{The system prompt for rewriting.}
    \label{fig:prompt6}
\end{figure*}

\begin{figure*}[t]
    \centering
    \includegraphics[width=\textwidth]{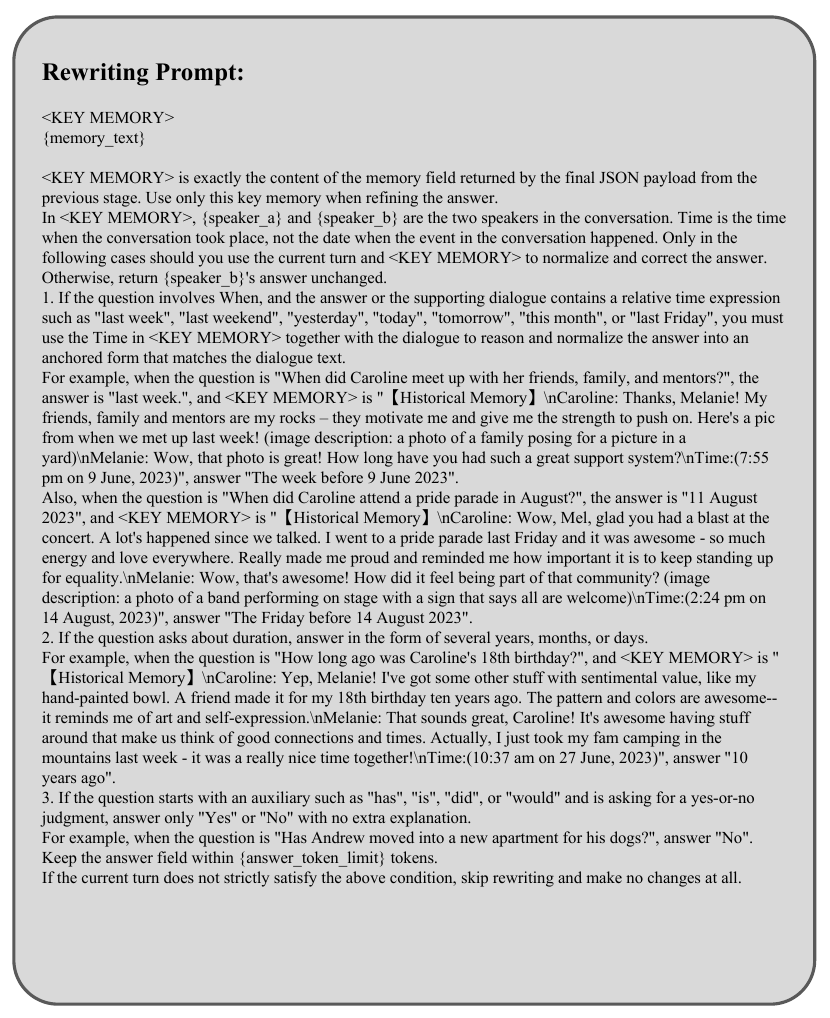}
    \caption{The user prompt for rewriting.}
    \label{fig:prompt7}
\end{figure*}

\begin{figure*}[t]
    \centering
    \includegraphics[width=\textwidth]{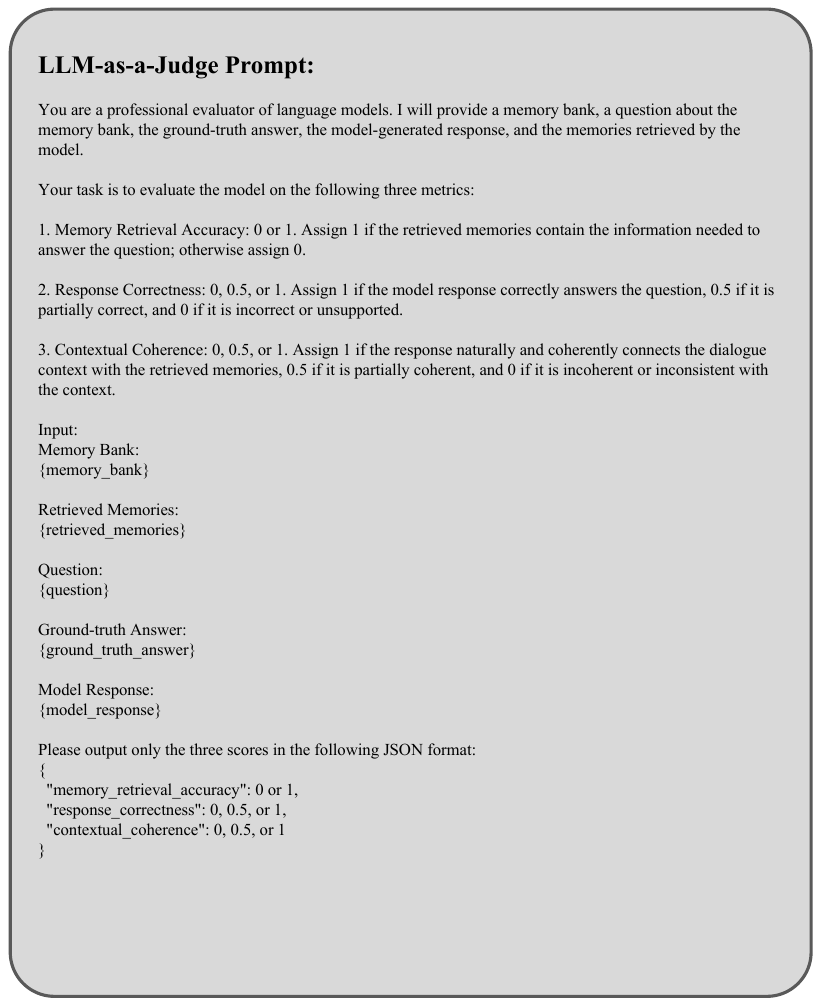}
    \caption{The LLM-as-a-Judge prompt for GVD evaluation.}
    \label{fig:prompt8}
\end{figure*}

We report all prompts used in {MGRetrieval}.
(1) Figure~\ref{fig:prompt1} shows the prompt for memory keyword extraction, (2) Figure~\ref{fig:prompt2} shows the prompt for memory keyword matching, (3) Figure~\ref{fig:prompt3} shows the prompt for question keyword selection, (4) Figures~\ref{fig:prompt4} and~\ref{fig:prompt5} show the system and user prompts for answering, (5) Figures~\ref{fig:prompt6} and~\ref{fig:prompt7} show the system and user prompts for rewriting, and (6) Figure~\ref{fig:prompt8} shows the LLM-as-a-Judge prompt.
\end{document}